\documentclass[12pt,onecolumn]{IEEEtran}

\usepackage{makecell}
\usepackage{array}
\usepackage{graphicx,amssymb,amsmath}
\usepackage{multicol}
\usepackage[noadjust]{cite}
\usepackage{setspace}
\usepackage{subfigure}
\usepackage{graphicx}
\usepackage{float}
\usepackage {url}
\usepackage{stfloats}
\usepackage{amsthm,pifont}
\usepackage{flushend}
\usepackage{cases,subeqnarray}
\usepackage{bm,multirow,bigstrut}
\usepackage{amsmath, amsthm, amssymb}
\usepackage{textcomp}
\usepackage{latexsym,bm}
\usepackage{booktabs}
\usepackage{xcolor}
\usepackage{mathtools}
\usepackage{dsfont}
\usepackage{extarrows}
\usepackage{epsfig}
\usepackage{epsfig}
\usepackage{algpseudocode}
\usepackage{algorithmicx,algorithm}

\usepackage{cite}
\usepackage{bm}
\usepackage{cleveref}
\usepackage{multicol}       
\usepackage{multirow}       
\usepackage{array}          
\usepackage{colortbl}
\usepackage{makecell}
\definecolor{crimson}{RGB}{192,0,0}         
\definecolor{navy}{RGB}{47,85,151}         
\usepackage{bbding}
\usepackage{graphicx}

\usepackage{booktabs}
\theoremstyle{plain}

\theoremstyle{plain}

\usepackage{amsmath}

\usepackage{threeparttable}
\usepackage{pdfsync}
\usepackage{flushend}

\IEEEoverridecommandlockouts

\begin{document}

\title{Hybrid-Field Channel Estimation for XL-MIMO Systems with Stochastic Gradient Pursuit Algorithm}
\author{Hao Lei, Jiayi Zhang,~\IEEEmembership{Senior Member,~IEEE}, Zhe Wang, Bo Ai,~\IEEEmembership{Fellow,~IEEE}, and \\ Derrick Wing Kwan Ng,~\IEEEmembership{Fellow,~IEEE}

\thanks{H. Lei, J. Zhang, and Z. Wang are with the School of Electronic and Information Engineering, Beijing Jiaotong University, Beijing 100044, China  (e-mail: haolei@bjtu.edu.cn; jiayizhang@bjtu.edu.cn; zhewang$\_$77@bjtu.edu.cn).

B. Ai is with the State Key Laboratory of Rail Traffic Control and Safety, Beijing Jiaotong University, Beijing 100044, China (e-mail: boai@bjtu.edu.cn).

D. W. K. Ng is with the School of Electrical Engineering and Telecommunications, University of New South Wales, Sydney, NSW 2052, Australia (e-mail: w.k.ng@unsw.edu.au).
%
%
%
}
}
\maketitle

\begin{abstract}
Extremely large-scale multiple-input multiple-output (XL-MIMO) is crucial for satisfying the high data rate requirements of the sixth-generation (6G) wireless networks.
In this context, ensuring accurate acquisition of channel state information (CSI) with low complexity becomes imperative. 
Moreover, deploying an extremely large antenna array at the base station (BS) might result in some scatterers being located in near-field, while others are situated in far-field, leading to a hybrid-field communication scenario.
To address these challenges, this paper introduces two stochastic gradient pursuit (SGP)-based schemes for the hybrid-field channel estimation in two scenarios.
For the first scenario in which the prior knowledge of the specific proportion of the number of near-field and far-field channel paths is known, the scheme can effectively leverage the angular-domain sparsity of the far-field channels and the polar-domain sparsity of the near-field channels such that the channel estimation in these two fields can be performed separately.
For the second scenario which the proportion is not available, we propose an off-grid SGP-based channel estimation scheme, which iterates through the values of the proportion parameter based on a criterion before performing the hybrid-field channel estimation.
We demonstrate numerically that both of the proposed channel estimation schemes achieve superior performance in terms of both estimation accuracy and achievable rates while enjoying lower computational complexity compared with existing schemes.
Additionally, we reveal that as the number of antennas at the UE increases, the normalized mean square error (NMSE) performances of the proposed schemes remain basically unchanged, while the NMSE performances of existing ones improve.
Remarkably, even in this scenario, the proposed schemes continue to outperform the existing ones.
\end{abstract}



\IEEEpeerreviewmaketitle

\section{Introduction}

The sixth-generation (6G) wireless communications promise to empower various emerging applications, e.g., holographic telepresence, extended reality, connected and autonomous vehicles, etc \cite{[6]}.
To realize these possibilities, extremely large-scale multiple-input multiple-output (XL-MIMO) has emerged as one of the most pivotal techniques \cite{[1]}-\cite{[29]}. 
In practice, deploying an exceptionally large number of antennas at the base station (BS) is the predominant method for implementing XL-MIMO systems \cite{[1]}, \cite{[37]}, \cite{[30]}. 
In this way, XL-MIMO systems can significantly enhance spectral efficiency (SE) by efficiently multiplexing multiple users on the same time-frequency resource \cite{[30]}-\cite{[31]}. 
Furthermore, the substantial beamforming gain resulting from large-scale antennas can provide enhanced spatial resolution \cite{[7]}, \cite{[32]}, and ensuring reliable communications and marked gains in achievable data rates \cite{[8]}, \cite{[40]}, \cite{[36]}.

Compared with conventional massive MIMO (mMIMO) systems, the deployment of an extremely large number of antennas in XL-MIMO systems 
introduces new electromagnetic (EM) characteristics that cannot be overlooked \cite{[1]}, \cite{[29]}. 
Specifically, the EM fields radiated from the antennas can be categorized into near-field and far-field. 
The Rayleigh distance, typically adopted to distinguish between near-field and far-field, can span distances ranging from hundreds of meters to even more \cite{[1]}, \cite{[9]}, \cite{[2]}.
This suggests that scatters may be partially located in the far-field of the BS and partially in the near-field of the BS, leading to a hybrid-field communication scenario within practical XL-MIMO systems.
More importantly, to achieve desired performance in XL-MIMO systems, precise acquisition of channel state information (CSI) is essential.
However, the intricate communication environment introduces significant challenges to channel estimation in XL-MIMO systems, especially in hybrid-field scenarios.

\subsection{Prior Works}
From an EM region perspective, existing channel estimation schemes for XL-MIMO systems can be classified into two categories, i.e., near-field channel estimation schemes \cite{[4]}-\cite{[16]} and hybrid-field channel estimation schemes \cite{[17]}-\cite{[23]}.
For near-field channel estimation, both the user equipment (UE) and scatters are in the near-field of the BS.
Given this near-field property, the XL-MIMO channel modeling should consider the spherical wave characteristics instead of the planar wave characteristics typically associated with the far-field, ensuring accurate characterization of near-field channels.
For example, a polar-domain representation of XL-MIMO channels was introduced, leveraging near-field array response vectors to capture both distance and angle information \cite{[4]}.
Specifically, by utilizing the polar-domain transform matrix proposed in \cite{[4]}, the XL-MIMO channel can be transformed into its equivalent polar-domain representation, which exhibits remarkable sparsity.
This observed sparsity has motivated the development of several compressed sensing (CS)-based near-field channel estimation schemes \cite{[2]}-\cite{[10]} 
 and deep learning (DL)-based strategies \cite{[11]}. 
Moreover, the spatial non-stationary property of XL-MIMO channels, which is another emergent EM characteristic, should also be considered due to the extremely large array aperture \cite{[12]}-\cite{[14]}. 
Furthermore, the authors in \cite{[15]} and \cite{[16]} considered not only user activity patterns but also the spatial non-stationary property for XL-MIMO systems based on bilinear Bayesian inference.

The second category of XL-MIMO channel estimation addresses the hybrid-field scenario, where some scatters are located in the far-field of the BS, while others are situated in the near-field of the BS.
In such scenarios, XL-MIMO channels usually comprise both near-field and far-field path components.
To address this, a hybrid-field channel model was introduced in prior work \cite{[17]}, which incorporated an adjustable parameter to tune the proportion of near-field and far-field paths in the model.
By leveraging the polar-domain sparsity of near-field paths and the angular-domain sparsity of far-field paths, a hybrid-field OMP channel estimation scheme was developed in \cite{[17]}.
The key idea of the hybrid-field OMP is to separately estimate the far-field path components in the angular domain and the near-field path components in the polar domain.
To enhance estimation accuracy, several channel estimation schemes were proposed based on the hybrid-field OMP in \cite{[17]}, such as the support detection OMP (SD-OMP) \cite{[18]}, the convolutional autoencoder (CAE)-based OMP (CAE-OMP) \cite{[28]}, and the recursive information distillation network (RIDNet)-based OMP (RIDNet-OMP) \cite{[27]}.
Furthermore, to address the need for prior knowledge of the proportion of the number of near-field and far-field paths, the authors in \cite{[19]} first determined this proportion by systematically traversing all possible values and then performed channel estimation.
On the other hand, different from segmenting near-field and far-field paths for separate estimation, the authors in \cite{[20]} and \cite{[24]} considered them as a whole and leveraged DL networks for hybrid-field channel estimation.
In addition, there are several channel estimation schemes that estimate the key parameters (i.e., channel gain, angle, and distance parameters) to reconstruct hybrid-field XL-MIMO channels \cite{[21]}-\cite{[23]}.

\subsection{Our Contributions}

\begin{table*}[t]
  \begin{center}
  \fontsize{8}{13}\selectfont
  \renewcommand{\arraystretch}{1}

    \caption{Comparison of relevant literature with this paper.}
      \label{table1}
 \begin{threeparttable}
    \begin{tabular}  {
    !{\vrule width1.2pt}  c
    !{\vrule width1.2pt}  c
    !{\vrule width1.2pt}  c
    !{\vrule width1.2pt}  c
    !{\vrule width1.2pt}  c
    !{\vrule width1.2pt}  c
    !{\vrule width1.2pt}  c
    !{\vrule width1.2pt}} 

      \Xhline{1.2pt}
         \bf Ref.
        &  \makecell{ \bf Hybrid-field }
        &  { \makecell{ \bf Multi-antenna \\ \bf UEs } }
        & \makecell{ \bf Off-grid  }
        & \makecell{ \bf Independent on the proportion of \\ \bf near-field and far-field paths   }
        & \makecell{ \bf Two distinct \\ \bf sensing matrices   }
        & \makecell{ \bf Deep learning   }\\
      \Xhline{1.2pt}

      \makecell{\cite{[10]}}
      &  \makecell{   \XSolidBrush }
      &  \makecell{ \XSolidBrush }
      & \makecell{ \Checkmark  }
      & \makecell{ $ \bf{-} $   }
      &  \makecell{ \XSolidBrush }
      &  \makecell{ \XSolidBrush } \\
      \hline

      \makecell{\cite{[17]}}
      &  \makecell{   \Checkmark }
      &  \makecell{ \XSolidBrush }
      & \makecell{ \XSolidBrush  }
      & \makecell{ \XSolidBrush   }
      &  \makecell{ \Checkmark }
      &  \makecell{ \XSolidBrush } \\
      \hline

      \makecell{\cite{[18]} }
      &  \makecell{   \Checkmark }
      &  \makecell{ \XSolidBrush }
      & \makecell{ \XSolidBrush  }
      & \makecell{ \XSolidBrush   }
      &  \makecell{ \Checkmark }
      &  \makecell{ \XSolidBrush } \\
      \hline

      \makecell{\cite{[28]}}
      &  \makecell{   \Checkmark }
      &  \makecell{ \XSolidBrush }
      & \makecell{ \Checkmark  }
      & \makecell{ \XSolidBrush   }
      &  \makecell{ \Checkmark }
      &  \makecell{ \Checkmark } \\
      \hline

      \makecell{\cite{[27]}}
      &  \makecell{   \Checkmark }
      &  \makecell{ \XSolidBrush }
      & \makecell{ \Checkmark  }
      &  \makecell{   \Checkmark }
      &  \makecell{ \XSolidBrush }
      & \makecell{ \Checkmark  } \\
      \hline

      \makecell{\cite{[19]}}
      &  \makecell{   \Checkmark }
      &  \makecell{ \XSolidBrush }
      & \makecell{ \XSolidBrush  }
      & \makecell{ \Checkmark   }
      &  \makecell{ \Checkmark }
      &  \makecell{ \XSolidBrush } \\
      \hline

      \makecell{The proposed on-grid\\ hybrid-field SGP}
      &  \makecell{   \Checkmark }
      &  \makecell{ \Checkmark }
      & \makecell{ \XSolidBrush  }
      & \makecell{ \XSolidBrush   }
      &  \makecell{ \Checkmark }
      &  \makecell{ \XSolidBrush } \\
      \hline

      \makecell{The proposed off-grid \\ hybrid-field SGP}
      &  \makecell{   \Checkmark }
      &  \makecell{ \Checkmark }
      & \makecell{ \Checkmark  }
      & \makecell{ \Checkmark   }
      &  \makecell{ \Checkmark }
      &  \makecell{ \XSolidBrush } \\
      \Xhline{1.2pt}

    \end{tabular}
  \end{threeparttable}
  \end{center}
\end{table*}

Despite various efforts have been devoted to unlock the potential of XL-MIMO systems, there are still various drawbacks in the existing results.
For instance, several near-field channel estimation schemes have focused solely on the characteristics of near-field channels, e.g., \cite{[4]}-\cite{[16]}. 
However, this approach can lead to a mismatch with the structural characteristics of practical hybrid-field channels.
Even worse, applying this approach in practice may result in degraded performance in terms of NMSE in emerging hybrid-field communication scenarios \cite{[17]}.
On the other hand, existing CS-based hybrid-field channel estimation schemes have demonstrated subpar performance in low signal-to-noise ratio (SNR) scenarios \cite{[17]}-\cite{[19]}.
Additionally, it is worth noting that the majority of hybrid-field channel estimation schemes have primarily focused on analyzing XL-MIMO systems in which the UE is only equipped with a single antenna \cite{[17]}-\cite{[24]}, \cite{[23]}. 
However, in practice, UEs are already equipped with multiple antennas to achieve higher SE performance and data rates.
Indeed, compared to the scenario of a single-antenna UE, the scenario of a multi-antenna UE introduces two very intuitive changes.
Firstly, the signal dimensions that need to be processed, which are already substantial due to the deployment of large-scale antennas at the BS, will further increase.
This, in turn, intensifies the challenges associated with achieving low-complexity channel estimation.
Secondly, the impact of increasing the number of antennas at the UE on the NMSE performance of channel estimation has been insufficiently explored, especially in the context of hybrid-field scenarios within XL-MIMO systems, e.g., \cite{[17]}-\cite{[24]}, \cite{[23]}. 
Consequently, it is imperative to conduct a comprehensive analysis of the influence of the multiple-antenna UE on the performance of hybrid-field channel estimation schemes in XL-MIMO systems.
Moreover, it should be noted that the authors of \cite{[17]}, \cite{[18]}, and \cite{[19]} considered the angles and distances as discrete points within the angular and polar domains (i.e., ``on-grid"), respectively.
Conversely, in practical systems, the angles and distances are typically distributed continuously, known as ``off-grid".
Addressing the resolution limitations inherent in these on-grid algorithms, the authors of \cite{[28]}, \cite{[27]}, \cite{[20]}-\cite{[21]} have turned towards DL-based algorithms for enhancing hybrid-field channel estimation.
While DL-based solutions offer significant NMSE performance advantages under specific conditions \cite{[28]}, \cite{[27]}, \cite{[20]}-\cite{[21]}, they also have significant limitations, such as their applicability to other scenarios and dependence on a large number of high-quality datasets.
Therefore, the applicability and effectiveness of traditional off-grid algorithms within hybrid-field XL-MIMO systems remain an area that warrants further investigation.

Motivated by the aforementioned studies, in this paper, we delve into a hybrid-field channel model with a multi-antenna UE. Then, two novel stochastic gradient pursuit (SGP)-based hybrid-field channel estimation schemes are proposed for different scenarios in XL-MIMO systems.
The comparisons of relevant literature with this paper are summarized in Table I.
Our major contributions are outlined as follows.
\begin{itemize}
  \item We consider a hybrid-field XL-MIMO system with a uniform linear array (ULA)-based BS and a multi-antenna UE, where the channel is characterized by the superposition of two components, i.e., the near-field and far-field channels. Subsequently, a novel on-grid hybrid-field SGP channel estimation scheme is proposed to estimate the far-field and near-field path components individually when the proportion of the number of near-field and far-field paths is known. More importantly, we reveal the impact of the number of antennas at the BS and UE on the NMSE performance of the proposed and existing schemes.
  \item To eliminate the reliance on prior knowledge of the proportion of the number of near-field and far-field paths, we further introduce a modified off-grid hybrid-field SGP channel estimation scheme. This novel approach does not rely on such prior knowledge, as required in existing works \cite{[17]}-\cite{[28]}. Instead, it employs an iterative process to systematically determine the proportion by examining all possible values. Moreover, the off-grid hybrid-field SGP employs an alternating minimization
      approach to refine channel parameters. This can effectively address the resolution constraints associated with the fixed grids in sensing matrices.
  \item We present numerical results to demonstrate the effectiveness of the proposed on-grid and off-grid hybrid-field SGP, particularly in low SNR scenarios. Moreover, in contrast to the least-squares (LS) process in the hybrid-field OMP \cite{[17]}, \cite{[19]}, our proposed on-grid and off-grid hybrid-field SGP adopt the least mean squares (LMS) process to approximate the performance of the minimum mean-squared error (MMSE) process.
      Therefore, the proposed schemes require lower complexity than that of the hybrid-field OMP, yet remain resilient against noise.
      In addition, the superior NMSE performance achieved by the proposed schemes results in higher achievable rates\footnote{ Simulation codes are provided to reproduce the results in this paper: https://github.com/BJTU-MIMO.}.
\end{itemize}

\subsection{Organization and Notation}

The rest of this paper is organized as follows. In Section \ref{System and Channel Model}, we first present the signal model and then introduce the characteristics and the sparsity of the hybrid-field channel.
Next, the proposed hybrid-field channel estimation schemes and their features, along with an analysis of their complexity, are presented in Section \ref{Hybrid-Field Channel Estimation}.
Then, Section \ref{Simulation Results} provides the numerical results and performance analysis.
Finally, conclusions are summarized in Section \ref{Conclusions}.

{\textbf{\textit{Notation}}:} We denote column vectors and matrices by boldface lowercase letters $\rm {\bf{a}}$ and boldface uppercase letters $\rm {\bf{A}}$, respectively.
Conjugate, transpose, conjugate transpose, and pseudoinverse are represented as $(\cdot)^{*}$, $(\cdot)^{T}$, $(\cdot)^{H}$, and $(\cdot)^{\dag}$ respectively.
The $ M \times N $ real-valued matrix and the $ M \times N $ complex-valued matrix are denoted by $\mathbb{R}^{ M \times N}$ and $\mathbb{C}^{ M \times N}$, respectively.
We represent the circularly symmetric complex-valued Gaussian distribution by $\mathcal{CN}(0,\sigma^2)$, where $ \sigma^2 $ denotes the variance.
The uniform distribution from $a$ to $b$ is denoted by $\mathcal{U} (a,b)$.
The Euclidean norm, the Kronecker product, and the $ N \times N $ identity matrix are denoted by $ \|\cdot\| $, $ \otimes $, and $ {\bf{I}}_N $.
We denote the $  N \times 1 $ vector with all elements being zero by ${{\bf{0}}_{N \times 1}}$.
$\mathbb{E} \{ \cdot \} $ and $  {\rm{tr}}(\cdot) $ are the expectation operator the trace operator, respectively.
The $i$-th row and $j$-th column of the matrix $\rm {\bf{A}}$ is denoted by $ {\rm {\bf{A}}}(i,j) $.
The $j$-th column of the matrix $\rm {\bf{A}}$ is represented as $  {\rm {\bf{A}}}(:,j) $.
The vectorization operation of the matrix ${\rm {\bf{A}}} \in \mathbb{C}^{ M \times N} $ is denoted as $  {\rm{vec}}\left( {{\bf{A}}} \right) = \left[ {{\bf{A}}}(:,1)^T , {{\bf{A}}}(:,2)^T, \cdots ,  {{\bf{A}}}(:,N)^T  \right]^T $.
We denote the round down operation by $ \lfloor \cdot \rfloor $.
The absolute value of a number or the determinant of a matrix is represented as $ |\cdot| $.
\section{System and Channel Model}\label{System and Channel Model}
\subsection{System Model}

\begin{figure}
  \centering
  \includegraphics[width=3in]{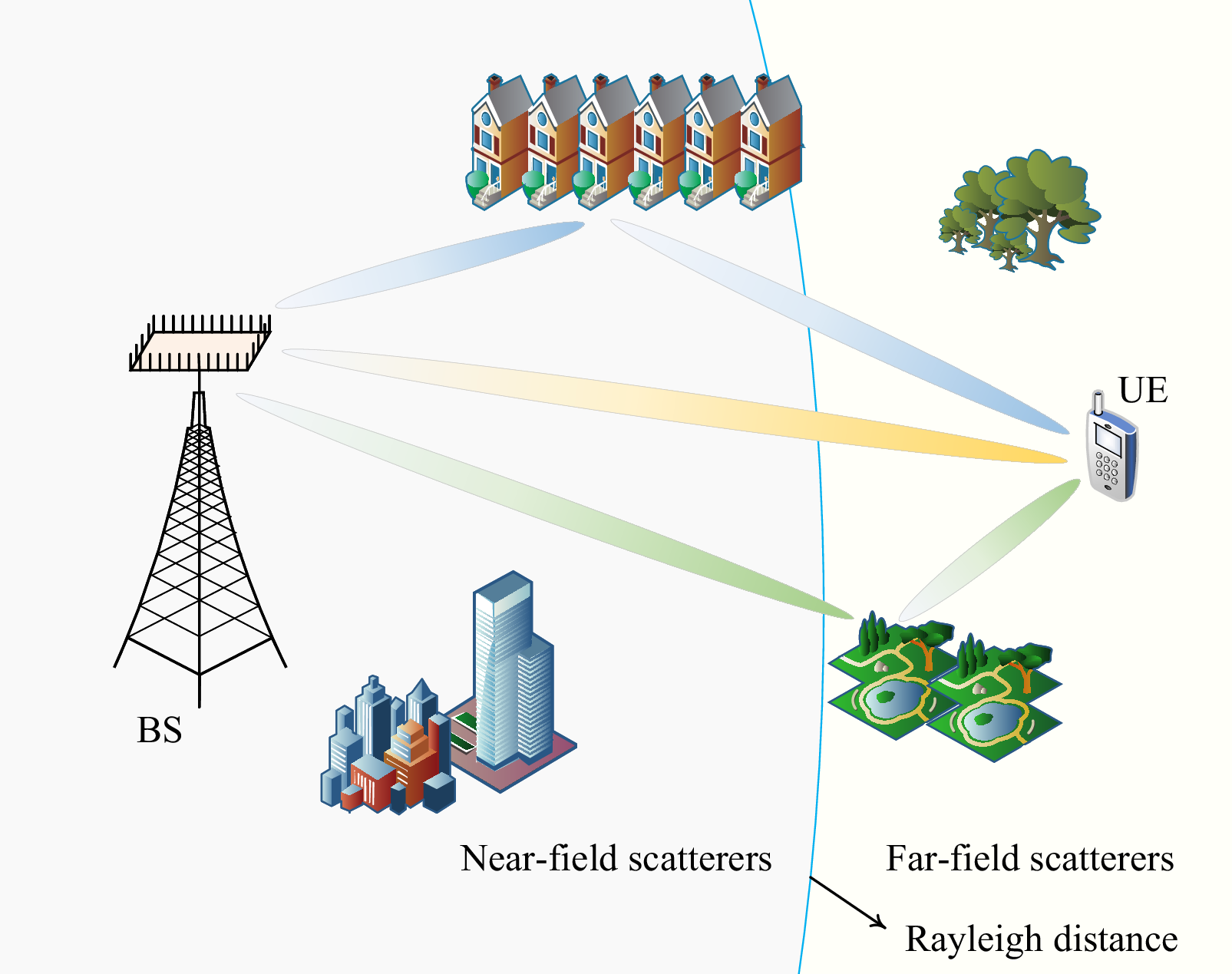}
  \caption{The hybrid-field communication scenario with a uniform linear array (ULA) at the BS in the XL-MIMO system, where the UE is equipped with multiple antennas. The Rayleigh distance is a widely adopted boundary to separate near-field and far-field communications.}
  \label{SYSTEM-MODEL}
\end{figure}

As illustrated in Fig. 1, an uplink time division duplexing (TDD) hybrid-field point-to-point communication scenario is considered in the XL-MIMO system, where the BS is equipped with a $ N_b $-element ULA and the UE\footnote{In the case of an XL-MIMO system with multiple UEs, the channels between the BS and its different UEs can be modeled and treated separately \cite{[30]}, \cite{[8]}. In addition, when multiple UEs transmit orthogonal pilots, channel estimation can be performed independently by linearly processing the received signal at the BS. Therefore, in this paper, we consider a typical UE equipped with multiple antennas in the considered XL-MIMO system to simplify the analysis and discussion.} is equipped with $ N_u $ antennas.
We denote the antenna spacing by $ \Delta = {\lambda  \mathord{\left/ {\vphantom {\lambda  2}} \right. \kern-\nulldelimiterspace} 2} $, where $ \lambda $ is the carrier wavelength.
Let $ {\bf{H}} \in {\mathbb{C}^{{N_b} \times {N_u}}} $ denote the channel between the BS and the UE.
When the UE transmit its pilot signal with length $ \tau $, the received signals, $ {\bf{Y}} \in {\mathbb{C}^{{N_b} \times \tau }} $, at the BS can be represented as
\begin{equation}\label{(1)}
  {\bf{Y}} = {\bf{H}}{\bf{P}}{\bf{\Phi}}^T +  {\bf{N}}    ,
\end{equation}
where $ {\bf{P}} \in {\mathbb{C}^{{N_u} \times N_u }}$ denotes the precoding matrix of the UE during the phase of pilot transmission,
$ \mathbf{\Phi}  \in {\mathbb{C}^{\tau  \times {N_u}}} $ is the pilot matrix sent by the UE to the BS such that $ {{\bf{\Phi}} ^T}{{\bf{\Phi}} ^*}= \tau {\bf{I}}_{N_u}$, $ {\bf{N}} \in {\mathbb{C}^{{N_b} \times \tau }}$ denotes the additive noise at the UE with independent identical distributed entries following $\mathcal{CN}(0,\sigma^2)$, and $ \sigma^2 $ is the noise power.
Note that the precoding matrix should satisfy the total power constraint $ {\rm{tr}}\left( {{\bf{P}}{{\bf{P}}^H}} \right) = p $, where $ p $ denotes the total transmit power of the UE.

To estimate $ {\bf{H}} $, the received signal is first correlated with the normalized pilot matrix $ {{\bf{\Phi}}^*  \mathord{\left/  {\vphantom { {\bf{\Phi}}  {\sqrt \tau  }}} \right. \kern-\nulldelimiterspace} {\sqrt \tau  }} $ to obtain $  {\bf{Y}}_t  = \frac{1}{{\sqrt \tau  }}{\bf{Y}}{{\bf{\Phi}} ^*} \in {\mathbb{C}^{{N_b} \times N_u }} $ at the BS, which is given by
\begin{equation}\label{(2)}
\begin{aligned}
    {\bf{Y}}_t &= \frac{1}{{\sqrt \tau  }}{\bf{H}} {\bf{P}} {{\bf{\Phi}} ^T}{{\bf{\Phi}} ^*} + \frac{1}{{\sqrt \tau  }}{\bf{N}}{{\bf{\Phi}} ^*} =\sqrt \tau  {\bf{H}}{\bf{P}} +  {\bf{N}}_t,
\end{aligned}
\end{equation}
where $ {\bf{N}}_t = \frac{1}{{\sqrt \tau  }} {\bf{N}}{{\bf{\Phi}} ^*} \in {\mathbb{C}^{{N_b} \times N_u }}$ is the equivalent noise.

\subsection{Channel Model}

It is interesting that the Rayleigh distance, denoted as $ d_{\rm Rayleigh} = {{2{D^2}} \mathord{\left/ {\vphantom {{2{D^2}} \lambda }} \right. \kern-\nulldelimiterspace} \lambda }$, can extend to hectometre range or longer in XL-MIMO systems, where $D$ represents the array aperture.
For instance, when the array aperture is $ 1 $ meters and the carrier wavelength is $ 0.01 $ meters, the Rayleigh distance can reach $200$ meters.
Due to the expanded near-field in XL-MIMO systems, it is pertinent to account for the possibility that the XL-MIMO channel might encompass both far-field and near-field components, as shown in Fig. 1.
Therefore, in a hybrid-field communication scenario, the XL-MIMO channel can be characterized by the superposition of two distinct components, i.e., the near-field channel and the far-field channel.
In addition, due to the spherical-wavefront in near-field and the planar-wavefront in far-field, their components should be modeled separately.

Without loss of generality, we assume that the XL-MIMO channel consists of one dominated line-of-sight (LoS) path and $L-1$ non-line-of-sight (NLoS) paths \cite{[3]}.
Then, based on the array response vectors \cite{[2]}, the hybrid-field XL-MIMO channel $ {\bf{H}} \in {\mathbb{C}^{{N_b} \times N_u }} $ can be modeled as
\begin{equation}\label{3}
\begin{aligned}
{\bf{H}} &= \sqrt {{N_b}{N_u}} {{\bar \alpha }_0}{\bf{a}}\left( {{r_{{\rm{R}},0}},{\theta _{{\rm{R}},0}}} \right){{\bf{a}}^H}\left( {{r_{{\rm{T}},0}},{\theta _{{\rm{T}},0}}} \right)\\
 &+ \sqrt {\frac{{{N_b}{N_u}}}{{L - 1}}} \sum\limits_{l = 1}^{L - 1} {{\alpha _l}{\bf{a}}\left( {{r_{{\rm{R}},l}},{\theta _{{\rm{R}},l}}} \right){{\bf{a}}^H}\left( {{r_{{\rm{T}},l}},{\theta _{{\rm{T}},l}}} \right)} \\
 & \mathop  = \limits^{\left( a \right)} \sqrt {\frac{{{N_b}{N_u}}}{{L - 1}}} \sum\limits_{l = 0}^{L - 1} {{\alpha _l}{\bf{a}}\left( {{r_{{\rm{R}},l}},{\theta _{{\rm{R}},l}}} \right){{\bf{a}}^H}\left( {{r_{{\rm{T}},l}},{\theta _{{\rm{T}},l}}} \right)},
\end{aligned}
\end{equation}
where $ L $\footnote{ {In practice, various techniques can be exploited to estimate the number of paths $L$ [35]. Thus, this paper assumes that $L$ is known in line with similar studies in [23]-[27] and centers on channel estimation given this knowledge, detailed in Section \uppercase\expandafter{\romannumeral3}.}} is the total number of the path components and $(a)$ holds by defining $ {\alpha _0} = \sqrt {L - 1} {{\bar \alpha }_0} $.
The array response vectors of the LoS path at the BS and the UE are $ {\bf{a}}\left( {{r_{{\rm{R}},0}},{\theta _{{\rm{R}},0}}} \right) \in {\mathbb{C}^{{N_b} \times 1 }}  $ and $ {\bf{a}}\left( {{r_{{\rm{T}},0}},{\theta _{{\rm{T}},0}}} \right) \in {\mathbb{C}^{{N_u} \times 1 }}  $, respectively.
The angle of the LoS path at the BS and the UE are $ {\theta _{{\rm{R}},0}} $ and $ {\theta _{{\rm{T}},0}} $, respectively.
We denote $ {r_{{\rm{R}},0}} = {r_{{\rm{T}},0}} $ as the distance between the BS and the UE.
The the path gain of the LoS path is denoted by $ {{\bar \alpha }_0} $ with $ {{\bar \alpha }_0}= \sqrt {{\kappa  \mathord{\left/ {\vphantom {\kappa  {( {1 + \kappa } )}}} \right. \kern-\nulldelimiterspace} {\left( {1 + \kappa } \right)}}}   $ \cite{[3]} while the gain of the $l$-th NLoS path is denoted by $ {\alpha _l} $ with $ {\alpha _l} \sim \mathcal{CN}(0,\sigma^2_l) $, where $ \sigma^2_l =  {1 \mathord{\left/ {\vphantom {1 {\left( {1 + \kappa } \right)}}} \right. \kern-\nulldelimiterspace} {\left( {1 + \kappa } \right)}}$, $\forall l\in \{1,2,\cdots,L-1\} $.
We denote the power ratio of the LoS and NLoS paths by the Rician factor $ \kappa \geq 0 $ \cite{[3]}.
The array response vectors of the NLoS paths at the BS and the UE are $ {\bf{a}}\left( {r_{{{\rm{R}},l}},\theta _{{\rm{R}},l}} \right) $ and $ {{\bf{a}}}\left( {r_{{\rm{T}},l}},{\theta _{{\rm{T}},l}} \right) $, $\forall l\in \{1,2,\cdots,L-1\} $, respectively.
More specifically, the array response vectors $ {\bf{a}}\left( {r_{{{\rm{R}},l}},\theta _{{\rm{R}},l}} \right) $ and $ {{\bf{a}}}\left( {r_{{\rm{T}},l}},{\theta _{{\rm{T}},l}} \right) $, $\forall l\in \{0,1,2,\cdots,L-1\} $, which are determined by the distances $  {r_{{\rm{R}},l}} $ and $  {r_{{\rm{T}},l}} $, and are given by
\begin{equation}\label{(4)}
\begin{aligned}
    {\bf{a}}\left( {r_{{{\rm{R}},l}},\theta _{{\rm{R}},l}} \right) &= \left\{ {\begin{array}{*{20}{l}}
{{\bf{a}}\left( {\theta _{{\rm{R}},l}} \right), \quad\quad\,\: {\rm{if}} \; {r_{{\rm{R}},l}} > d_{\rm Rayleigh},}\\
{{\bf{a}}\left( {r_{{{\rm{R}},l}},\theta _{{\rm{R}},l}} \right), \;\; {\rm{otherwise,}}}
\end{array}} \right. \\
   {{\bf{a}}}\left( {r_{{\rm{T}},l}},{\theta _{{\rm{T}},l}} \right) &= \left\{ {\begin{array}{*{20}{l}}
{{\bf{a}}\left( {\theta _{{\rm{T}},l}} \right), \quad\quad\,\: {\rm{if}} \; {r_{{\rm{T}},l}} > d_{\rm Rayleigh},}\\
{{{\bf{a}}}\left( {r_{{\rm{T}},l}},{\theta _{{\rm{T}},l}} \right), \;\; {\rm{otherwise}},}
\end{array}} \right.
\end{aligned}
\end{equation}
respectively. To further unveil the features of the hybrid-field channel in XL-MIMO, the far-field and the near-field components are discussed in details as follows:

\subsubsection{Far-field Channel}

We assume that there are $L_{\rm{F}} = \lfloor \gamma L \rfloor$ far-field components among all the $L$ paths, where $ 0 \leq \gamma \leq 1 $ is an adjustable parameter.
Then, the far-field channel $ {{\bf{H}}_{{\rm{far - field}}}} \in {\mathbb{C}^{{N_b} \times N_u }} $ can be modeled by the planar-wavefront, which can be expressed as \cite{[4]}, \cite{[3]}
\begin{equation}\label{(5)}
\begin{aligned}
{{\bf{H}}_{{\rm{far - field}}}}  =  \sqrt {\frac{{{N_b}{N_u}}}{L-1}} \sum\limits_{{l_{\rm{f}}} = 1}^{{L_{\rm{F}}}} {{\alpha _{{l_f}}}{\bf{a}}\left( {\theta _{{\rm{R}},{l_{\rm{f}}}}} \right){{\bf{a}}^H}\left( {\theta _{{\rm{T}},{l_{\rm{f}}}}} \right)}.
\end{aligned}
\end{equation}
Based on the planar-wavefront assumption, the far-field array response vectors are represented as \cite{[4]}, \cite{[3]}
\begin{equation}\label{(6)}
\begin{aligned}
{\bf{a}}\left( {\theta _{{\rm{R}},{l_{\rm{f}}}}} \right) = \frac{1}{{\sqrt {{N_b}} }}{\left[ {1,{e^{ - j\pi {\theta _{{\rm{R}},{l_{\rm{f}}}}}}}, \cdots ,{e^{ - j(N_b - 1)\pi {\theta _{{\rm{R}},{l_{\rm{f}}}}}}}} \right]^H},\\
{{\bf{a}}}\left( {\theta _{{\rm{T}},{l_{\rm{f}}}}} \right) = \frac{1}{{\sqrt {{N_u}} }}{\left[ {1,{e^{ - j\pi {\theta _{{\rm{T}},{l_{\rm{f}}}}}}}, \cdots ,{e^{ - j(N_u - 1)\pi {\theta _{{\rm{T}},{l_{\rm{f}}}}}}}} \right]^H},
\end{aligned}
\end{equation}
respectively. where $ {\theta _{{\rm{R}},{l_{\rm{f}}}}} $ and $ {\theta _{{\rm{T}},{l_{\rm{f}}}}}$ are the angle of the $ {{l_{\rm{f}}} } $-th far-field path at the BS and the UE, respectively.
Note that $ {\theta _{{\rm{R}},{l_{\rm{f}}}}} = \sin (\upsilon_{{\rm{R}},{l_{\rm{f}}}}) $ and $ {\theta _{{\rm{T}},{l_{\rm{f}}}}} = \sin (\upsilon_{{\rm{T}},{l_{\rm{f}}}}) $, where $ \upsilon_{{\rm{R}},{l_{\rm{f}}}} \in ( - {\pi  \mathord{\left/{\vphantom {\pi  2}} \right. \kern-\nulldelimiterspace} 2},  {\pi  \mathord{\left/{\vphantom {\pi  2}} \right. \kern-\nulldelimiterspace} 2} ) $ and $ \upsilon_{{\rm{T}},{l_{\rm{f}}}} \in ( - {\pi  \mathord{\left/{\vphantom {\pi  2}} \right. \kern-\nulldelimiterspace} 2},  {\pi  \mathord{\left/{\vphantom {\pi  2}} \right. \kern-\nulldelimiterspace} 2} ) $ are the practical physical angles at the BS and the UE, respectively.

It is worth noting that the corresponding angular-domain representation $ {{\bf{H}}_{{\rm{far - field}}}^{\rm A}} \in {\mathbb{C}^{{N_b} \times N_u }} $ can be derived from the channel $ {{\bf{H}}_{{\rm{far - field}}}} $ based on the pair of discrete Fourier transform (DFT) matrices $  {\bf{F}}_{\rm R} $ and $  {\bf{F}}_{\rm T} $, which is given by
\begin{equation}\label{(7)}
\begin{aligned}
{{\bf{H}}_{{\rm{far - field}}}} = {\bf{F}}_{\rm R}  {{\bf{H}}_{{\rm{far - field}}}^{\rm A}} {\bf{F}}_{\rm T}^H,
\end{aligned}
\end{equation}
where $ {\bf{F}}_{\rm R} =   \left[ {{\bf{a}}\left( {{\theta _1}} \right), \cdots ,{\bf{a}}\left( {{\theta _{{n_b}}}} \right), \cdots ,{\bf{a}}\left( {{\theta _{{N_b}}}} \right)} \right]  \in {\mathbb{C}^{{N_b} \times N_b }}   $ and $ {\bf{F}}_{\rm T} =   \left[ {{\bf{a}}\left( {{\theta _1}} \right), \cdots ,{\bf{a}}\left( {{\theta _{{n_u}}}} \right), \cdots ,{\bf{a}}\left( {{\theta _{{N_u}}}} \right)} \right] \in {\mathbb{C}^{{N_u} \times N_u }} $ are unitary matrices, $ {\theta _{{n_b}}} = \frac{{2{n_b} - {N_b} - 1}}{{{N_b}}} $ with $ {n_b} = 1,2, \cdots ,{N_b} $, and $ {\theta _{{n_u}}} = \frac{{2{n_u} - {N_u} - 1}}{{{N_u}}} $ with $ {n_u} = 1,2, \cdots ,{N_u} $.
Due to the limited scatterers (i.e., $ {{L_{\rm{F}}}} \ll N_b $) \cite{[34]}, \cite{[35]}, the angular-domain representation $ {{\bf{H}}_{{\rm{far - field}}}^{\rm A}} $ displays remarkable sparsity.
Indeed, this sparsity has been exploited by several CS-based channel estimation schemes to effectively address far-field channel estimation problems \cite{[34]}, \cite{[35]}.

\subsubsection{Near-field Channel}
For modeling the near-field channel, in contrast to the planar-wavefront assumption in the far-field, the near-field channel should be modeled by a spherical-wavefront.
This means that the far-field array response vectors must be substituted with the near-field array response vectors to more accurately model the near-field channel\footnote{{{It is worth noting that the far-field channel is a special case of the near-field channel. When the distance between the UE and the BS is greater than the Rayleigh distance, the near-field channel will degenerate into a far-field channel. Besides, it is worth using an accurate (but also complex) model in terms of attainable performance. There are already numerous studies suggesting that leveraging more accurate but more complex near-field channel models can achieve better performance [6]-[11].}}}.
As such, the near-field channel can be expressed as \cite{[2]}, \cite{[4]}
\begin{equation}\label{(8)}
\begin{aligned}
{{\bf{H}}_{{\rm{near - field}}}} \!\! = \!\!  \sqrt {\frac{{{N_b}{N_u}}}{L-1}} \sum\limits_{{l_{\rm{n}}} = 1}^{{L_{\rm{N}}}} \!\!{{\alpha _{{l_{\rm{n}}}}}{\bf{a}}\left( {r_{{\rm{R}},{l_{\rm{n}}}},\theta _{{\rm{R}},{l_{\rm{n}}}}} \right){{\bf{a}}^H}\!\left( {r_{{\rm{T}},{l_{\rm{n}}}},\theta _{{\rm{T}},{l_{\rm{n}}}}} \right)},
\end{aligned}
\end{equation}
where $ {L_{\rm{N}}} = L - {L_{\rm{F}}}$ is the number of near-field components among all the $ L $ paths. 
We denote the near-field array response vectors at the BS and the UE by $ {\bf{a}}\left( {r_{{\rm{R}},{l_{\rm{n}}}},\theta _{{\rm{R}},{l_{\rm{n}}}}} \right) $ and $ {{\bf{a}}}\left( {r_{{\rm{T}},{l_{\rm{n}}}},\theta _{{\rm{T}},{l_{\rm{n}}}}} \right) $, respectively, which can be denoted by \cite{[2]}
\begin{equation}\label{(9)}
\begin{aligned}
  {\bf{a}}\left( {r_{{\rm{R}},{l_{\rm{n}}}},\theta _{{\rm{R}},{l_{\rm{n}}}}} \right) = &\frac{1}{{\sqrt {{N_b}} }}[ {{e^{ - j\frac{{2\pi }}{\lambda }( {r_{{\rm{R}},{l_{\rm{n}}}}^{\left( 1 \right)} - r_{{\rm{R}},{l_{\rm{n}}}}^{}} )}}, \cdots , }\\& {e^{ - j\frac{{2\pi }}{\lambda }( {r_{{\rm{R}},{l_{\rm{n}}}}^{\left( {{N_b}} \right)} - r_{{\rm{R}},{l_{\rm{n}}}}^{}} )}} ]^T,\\
  {{\bf{a}}}\left( {r_{{\rm{T}},{l_{\rm{n}}}},\theta _{{\rm{T}},{l_{\rm{n}}}}} \right) =& \frac{1}{{\sqrt {{N_u}} }}{ [ {{e^{ - j\frac{{2\pi }}{\lambda }( {r_{{\rm{T}},{l_{\rm{n}}}}^{\left( 1 \right)} - r_{{\rm{T}},{l_{\rm{n}}}}^{}} )}}, \cdots ,}} \\ & {{ {e^{ - j\frac{{2\pi }}{\lambda }( {r_{{\rm{T}},{l_{\rm{n}}}}^{\left( {{N_u}} \right)} - r_{{\rm{T}},{l_{\rm{n}}}}^{}} )}}} ]^T},
\end{aligned}
\end{equation}
respectively, where $ \theta _{{\rm{R}},{l_{\rm{n}}}} $ ($ \theta _{{\rm{T}},{l_{\rm{n}}}} $) denotes the angle of the $ {l_{\rm{n}}} $-th near-field path at the BS (UE) and $ r_{{\rm{R}},{l_{\rm{n}}}} $ ($ r_{{\rm{T}},{l_{\rm{n}}}} $) denotes the distance between the $ {l_{\rm{n}}} $-th scatterer and the center of the antenna array of the BS (UE).
We denote the distance between the $ {l_{\rm{n}}} $-th scatterer and the $ n_b $-th antenna of the BS by $ r_{{\rm{R}},{l_{\rm{n}}}}^{{n_b}} = \sqrt {{{\left( {{r_{{\rm{R}},{l_{\rm{n}}}}}} \right)}^2} + \omega _{{n_b}}^2{\Delta ^2} - 2{r_{{\rm{R}},{l_{\rm{n}}}}}\omega _{{n_b}}^{}\Delta \sin \left( {\theta _{{\rm{R,}}{l_{\rm{n}}}}^{}} \right)}  $ with $ \omega _{{n_b}}^{} = \frac{{2{n_b} - {N_b} - 1}}{2} $, $ {n_b} = 1,2, \cdots ,{N_b} $.
$ r_{{\rm{T}},{l_{\rm{n}}}}^{{n_u}} = \sqrt {{{\left( {{r_{{\rm{T}},{l_{\rm{n}}}}}} \right)}^2} + \omega _{{n_u}}^2{\Delta ^2} - 2{r_{T,{l_{\rm{n}}}}}\omega _{{n_u}}^{}\Delta \sin \left( {\theta _{{\rm{T,}}{l_{\rm{n}}}}^{}} \right)}  $ denotes the distance between the $ {l_{\rm{n}}} $-th scatterer and the $ n_u $-th antenna of the UE with $ \omega _{{n_u}}^{} = \frac{{2{n_u} - {N_u} - 1}}{2} $, $ {n_u} = 1,2, \cdots ,{N_u} $.

Note that far-field channels are well-known for exhibiting significant sparsity in the angular domain, while near-field channels demonstrate pronounced sparsity in the polar domain \cite{[2]}, \cite{[4]}. 
To transform the near-field channel to its polar-domain representation, we first review the polar-domain transform matrices, which can be expressed as \cite{[4]}
\begin{equation}\label{(10)}
\begin{aligned}
{{\bf{D}}_{\rm{R}}} =& [{\bf{a}}({r_{{\theta _1},1}},{\theta _1}), \cdots ,{\bf{a}}({r_{{\theta _1},{m_1}}},{\theta _1}), \cdots ,\\
&{\bf{a}}({r_{{\theta _1},{M_1}}},{\theta _1}), \cdots ,{\bf{a}}({r_{{\theta _{{N_b}}},1}},{\theta _{{N_b}}}), \cdots ,\\
&{\bf{a}}({r_{{\theta _{{N_b}}},{m_{{N_b}}}}},{\theta _{{N_b}}}), \cdots ,{\bf{a}}({r_{{\theta _{{N_b}}},{M_{{N_b}}}}},{\theta _{{N_b}}})],\\
{{\bf{D}}_{\rm{T}}} =& [{\bf{a}}({r_{{\theta _1},1}},{\theta _1}), \cdots ,{\bf{a}}({r_{{\theta _1},{m_1}}},{\theta _1}), \cdots ,\\
&{\bf{a}}({r_{{\theta _1},{M_1}}},{\theta _1}), \cdots ,{\bf{a}}({r_{{\theta _{{N_u}}},1}},{\theta _{{N_u}}}), \cdots ,\\
&{\bf{a}}({r_{{\theta _{{N_u}}},{m_{{N_u}}}}},{\theta _{{N_u}}}), \cdots ,{\bf{a}}({r_{{\theta _{{N_u}}},{M_{{N_u}}}}},{\theta _{{N_u}}})],
\end{aligned}
\end{equation}
respectively, where the matrices $ {{\bf{D}}_{\rm{R}}} \in {\mathbb{C}^{{N_b} \times M_b }} $ and $ {{\bf{D}}_{\rm{T}}} \in {\mathbb{C}^{{N_u} \times M_u }} $ are the polar-domain transform matrices at the BS and the UE, respectively.
Each column of the matrix $ {{\bf{D}}_{\rm{R}}} $ ($ {{\bf{D}}_{\rm{T}}} $) is an array response vector with the distance $ {{r_{{\theta _{{n_b}}},{m_{{n_b}}}}}} $ ($ {{r_{{\theta _{{n_u}}},{m_{{n_u}}}}}} $) and the angle $ {{\theta _{{n_b}}}} $ ($ {{\theta _{{n_u}}}} $), where $ {m_{{n_b}}} \in [1,2, \cdots ,{M_{{n_b}}}] $ ($ {m_{{n_u}}} \in [1,2, \cdots ,{M_{{n_u}}}] $) and $ {n_b} \in [1,2, \cdots ,{N_b}] $ ($ {n_u} \in [1,2, \cdots ,{N_u}] $).
We denote the number of sampled distances at the sampled angle $ {{\theta _{{n_b}}}} $ ($ {{\theta _{{n_u}}}} $) by $ M_{n_b} $ ($ M_{n_u} $).
Thus, the numbers of all the sampled distances at the BS and the UE is denoted by $ {M_b} = \sum\nolimits_{{n_b} = 1}^{{N_b}} {{M_{{n_b}}}}  $ and $ {M_u} = \sum\nolimits_{{n_u} = 1}^{{N_u}} {{M_{{n_u}}}} $, respectively.

Based on the matrices $ {{\bf{D}}_{\rm{R}}}  $ and $ {{\bf{D}}_{\rm{T}}}  $, the corresponding polar-domain representation $ {{\bf{H}}_{{\rm{near - field}}}^{\rm P }} \in {\mathbb{C}^{{M_b} \times M_u }} $ is given by
\begin{equation}\label{(11)}
\begin{aligned}
{{\bf{H}}_{{\rm{near - field}}}} = {{\bf{D}}_{\rm{R}}}{\bf{H}}_{{\rm{near - field}}}^{\rm{P}}{\bf{D}}_{\rm{T}}^H.
\end{aligned}
\end{equation}

Then, the hybrid-field XL-MIMO channel $ {\bf{H}}  $ can be represented as
\begin{equation}\label{(12)}
\begin{aligned}
{\bf{H}} &= {{\bf{H}}_{{\rm{far - field}}}} + {{\bf{H}}_{{\rm{near - field}}}}\\
&={\bf{F}}_{\rm R}  {{\bf{H}}_{{\rm{far - field}}}^{\rm A}} {\bf{F}}_{\rm T}^H + {{\bf{D}}_{\rm{R}}}{\bf{H}}_{{\rm{near - field}}}^{\rm{P}}{\bf{D}}_{\rm{T}}^H.
\end{aligned}
\end{equation}
From \eqref{(12)}, it can be seen that the hybrid-field channel is composed of a combination of far-field and near-field channels. Therefore, if the specific number of the channel paths in far-field and near-field can be determined, it becomes possible to estimate the far-field and near-field channels separately.
Subsequently, these estimated channels can be utilized to reconstruct the whole hybrid-field channel.

\subsection{Uplink Data Transmission}

When the UE transmits its information signal $ {\bf{x}} = {\left[ {{x_1}, \cdots ,{x_{{N_u}}}} \right]^T} \in {\mathbb{C}^{{N_u} \times 1 }} $ to the BS, the received signal $ {\bf{y}} \in {\mathbb{C}^{{N_b} \times 1 }} $ at the BS is
\begin{equation}\label{(112)}
\begin{aligned}
{\bf{y}} = {\bf{H}}{{\bf{P}}_u}{\bf{x}} + {\bf{n}}_u,
\end{aligned}
\end{equation}
where $ {\bf{x}}  \sim  \mathcal{CN}(0, {{\bf{I}}_{{N_u}}} )    $, $ {\bf{P}}_u \in {\mathbb{C}^{{N_u} \times N_u }}$ denotes the precoding matrix of the UE during the phase of uplink data transmission, $ {\bf{n}}_u  \sim  \mathcal{CN}(0, \sigma_u^2 {{\bf{I}}_{{N_b}}} )  $ represents the independent receiver noise.
Note that the matrix $ {\bf{P}}_u $ satisfies the power constraint $ {\rm{tr}}\left( {{\bf{P}}_u {{\bf{P}}_u^H}} \right) = p $, where $ p $ denotes the total transmit power of the UE.

Let $ {\bf{V}} \in {\mathbb{C}^{{N_b} \times N_u }} $ represent the receive combining matrix selected by the BS for the UE.
Then, the estimate of $ {\bf{x}} $ can be given by
\begin{equation}\label{(113)}
\begin{aligned}
{\bf{\hat x}} = {\bf{ V}}^H  {\bf{ H}} {{\bf{P}}_u}{\bf{x}} + {\bf{ V}}^H {\bf{ n}}.
\end{aligned}
\end{equation}

Based on \eqref{(113)}, the achievable rate\footnote{{{It is important to recognize that the DoF in a system are fundamentally dictated by the channel characteristics and serve as an indicator of the system's potential performance ceiling [2], [8]. Nevertheless, in practical signal processing and performance evaluation, reliance on channel estimation data (which is typically imperfect CSI) is inevitable [38]. Therefore, achieving precise channel estimation is crucial for tapping into the maximum achievable rate that the system can offer.}}} for the UE can be expressed as \cite{[8]}, \cite{[38]}
\begin{equation}\label{(114)}
\begin{aligned}
R = \mathbb{E} \left\{ {{{\log }_2}\left| {{{\bf{I}}_{{N_u}}} + {{\bf{\Sigma }}^H}{ {\bf \Xi} ^{ - 1}}{\bf{\Sigma }}} \right| } \right\},  
\end{aligned}
\end{equation}
where $ {{\bf{\Sigma }}} = {\bf{ V}}^H {\bf{H}}{{\bf{P}}_u}  $, ${\bf \Xi } = \sigma_u^2  {\bf{ V}}^H{\bf{ V}} $, and the expectation is with respect to the channel estimates.
If maximum ratio (MR) combining $ {\bf{ V}} = {\bf{\hat H}} $ is employed, the achievable rate in \eqref{(114)} can be reformulated as
\begin{equation}\label{(115)}
\begin{aligned}
R =  \mathbb{E} \left\{ {{{\log }_2}\left|{   {{\bf{I}}_{{N_u}}} +    {{\bf{P}}_u^H}{{\bf{H}}^H}{\bf{\hat H}}   \left(   {\sigma_u^2  {\bf{ V}}^H{\bf{ V}}}\right) ^{ - 1}        {\bf{\hat H}}^H {\bf{H}}{{\bf{P}}_u}      }\right|}  \right\} . %
\end{aligned}
\end{equation}
From \eqref{(115)}, it can be seen that the quality of channel estimation can directly affect the achievable rate, which will be discussed in Section \ref{The SE performance}.

\section{Hybrid-Field Channel Estimation}\label{Hybrid-Field Channel Estimation}

Before introducing the hybrid-field channel estimation, we first rewrite the received pilot signal $ {\bf{Y}}_t $ to exploit the angular-domain sparsity for far-field and the polar-domain sparsity for near-field by plugging \eqref{(12)} into \eqref{(2)}, which can be represented as
\begin{equation}\label{(13)}
\begin{aligned}
{{\bf{Y}}_t} &= \sqrt \tau  {{\bf{F}}_{\rm{R}}}{\bf{H}}_{{\rm{far - field}}}^{\rm{A}}{\bf{F}}_{\rm{T}}^H{\bf{P}}  \\
  & + \sqrt \tau  {{\bf{D}}_{\rm{R}}}{\bf{H}}_{{\rm{near - field}}}^{\rm{P}}{\bf{D}}_{\rm{T}}^H{\bf{P}} + {{\bf{N}}_t}.
\end{aligned}
\end{equation}
According to $ {\rm{vec}}\left( {{\bf{ABC}}} \right) = \left( {{{\bf{C}}^T} \otimes {\bf{A}}} \right){\rm{vec}}\left( {\bf{B}} \right) $ \cite{[26]}, the matrix $ {{\bf{Y}}_t} $ can be reformulated as
\begin{equation}\label{(14)}
\begin{aligned}
{{\bf{y}}_t} &= \sqrt \tau  \left( {{{\bf{P}}^T}{\bf{F}}_{\rm{T}}^* \otimes {{\bf{F}}_{\rm{R}}}} \right){\bf{h}}_{{\rm{far - field}}}^{\rm{A}}  \\& + \sqrt \tau  \left( {{{\bf{P}}^T}{\bf{D}}_{\rm{T}}^* \otimes {{\bf{D}}_{\rm{R}}}} \right){\bf{h}}_{{\rm{near - field}}}^{\rm{P}} + {{\bf{n}}_t},
\end{aligned}
\end{equation}
where $ {{\bf{y}}_t} = {\rm{vec}}\left( {{{\bf{Y}}_t}} \right)  \in {\mathbb C} {^{{N_b}{N_u} \times 1}} $, $ {\bf{h}}_{{\rm{far - field}}}^{\rm{A}} = {\rm{vec}}\left( {{\bf{H}}_{{\rm{far - field}}}^{\rm{A}}} \right) \in {\mathbb C} {^{{N_b}{N_u} \times 1}}$, $ {\bf{h}}_{{\rm{near - field}}}^{\rm{P}} = {\rm{vec}}\left( {{\bf{H}}_{{\rm{near - field}}}^{\rm{P}}} \right)\in {\mathbb C} {^{{M_b}{M_u} \times 1}} $, and $ {{\bf{n}}_t} = {\rm{vec}}\left( {{{\bf{N}}_t}} \right) \in {\mathbb C} {^{{N_b}{N_u} \times 1}}  $.
As discussed above, both the near-field polar-domain channel $ {\bf{h}}_{{\rm{near - field}}}^{\rm{P}} $ and the far-field angular-domain channel $ {\bf{h}}_{{\rm{far - field}}}^{\rm{A}} $ exhibit remarkable sparsity.
Therefore, the channel estimation problem can be transformed into a sparse reconstruction problem and then several CS algorithms can be utilized to solve this problem.
For instance, the OMP algorithm was exploited to perform hybrid-field channel estimation by estimating the near-field and far-field path components, respectively, with the transform matrices \cite{[17]}.
Specifically, the OMP algorithm consists of two core procedures, i.e., the pursuing process and the LS process.
However, due to the employment of the LS algorithm, the performance of the signal reconstruction for the OMP is generally susceptible to noise \cite{[25]}.
Therefore, the hybrid-field OMP channel estimation scheme cannot achieve satisfactory estimation accuracy in low SNR scenarios.

To address this, the authors in \cite{[25]} proposed a SGP algorithm by substituting the LS process with the LMS process, which is a stochastic gradient approach to iteratively approximate the MMSE process.
Inspired by \cite{[17]} and \cite{[25]}, the SGP algorithm is exploited to perform the hybrid-field channel estimation by estimating the near-field and the far-field channels separately.
In addition, the proposed hybrid-field SGP channel estimation scheme are expected to exhibit robustness against noise.
In the following subsections, we will delve into our proposed hybrid-field channel estimation schemes from the perspective of whether the scheme is independent on the prior knowledge of $ \gamma $ or not.

\begin{algorithm}[t]
  \caption{The Proposed On-grid Hybrid-field SGP Channel Estimation Scheme with Prior Knowledge $\gamma$. }
  \label{alg:1}
  \begin{algorithmic}[1]
  \Require ${\bf{y}}_t$, ${\bf{P}}$, ${\bf{F}}_{\rm{T}}$, ${\bf{F}}_{\rm{R}}$, ${\bf{D}}_{\rm{T}}$, ${\bf{D}}_{\rm{R}}$, $L$, $\tau$, $ \gamma $, $\mu$
  \Ensure ${\bf{\hat H}} $
  {\Statex  \textbf{Initialization:} $L_{\rm{F}} = \gamma L$, $ L_{\rm{N}} = (1-\gamma)L $, $ \Omega_{\rm{F}}=\Omega_{\rm{N}} = \emptyset $, $ {\bf{r}} = {\bf{y}}_t $}.
  \Statex // \textbf{Phase \uppercase\expandafter{\romannumeral1}:} Estimating the far-field path components in the angular domain.

    \State $ {\bf{A}}_{\rm{F}} =  \sqrt \tau  \left( {{{\bf{P}}^T}{\bf{F}}_{\rm{T}}^* \otimes {{\bf{F}}_{\rm{R}}}} \right) $, $ {\bf{\hat h}}_{{\rm{far - field}}}^{\rm{A}} = {{\bf{0}}_{N_b N_u \times 1}}$.

    \For {$ l_f = 1,2,\cdots,L_{\rm{F}}$}
        \State Update the support set: $  \eta = \mathop {\arg \max }\limits_{n = 1,2, \cdots ,N} \left\| {{\bf{A}}_{\rm{F}}^H\left( {:,n} \right){\bf{r}}} \right\|_2^2   $, $ \Omega_{\rm{F}} = \Omega_{\rm{F}} \bigcup \eta $.
        \State Perform the LMS process: ${\bf{x}} = {\bf{\hat h}}_{{\rm{far - field}}}^{\rm{A}}(\Omega_{\rm{F}})$,
        \For{$ n = 1,2,\cdots,N_b N_u$}
            \State ${\bf{c}} = {\bf{A}}_f(n,\Omega_{\rm{F}})$, $ d = {\bf{y}}(n) $, $ e = d - {\bf{c}} {\bf{x}} $,
            \State $ {\bf{x}} = {\bf{x}} + \mu e {\bf{c}}^H $,
        \EndFor
        \State ${\bf{\hat h}}_{{\rm{far - field}}}^{\rm{A}}(\Omega_{\rm{F}}) = {\bf{x}}$.
        \State Compute the residual vector: $ {\bf{r}} = {\bf{y}} - {\bf{A}}_{\rm{F}}{\bf{\hat h}}_{{\rm{far - field}}}^{\rm{A}} $.
    \EndFor

  \Statex // \textbf{Phase \uppercase\expandafter{\romannumeral2}:} Estimating the near-field path components in the polar domain.
    \State $ {\bf{A}}_{\rm{N}} =  \sqrt \tau  \left( {{{\bf{P}}^T}{\bf{D}}_{\rm{T}}^* \otimes {{\bf{D}}_{\rm{R}}}} \right) $, ${\bf{\hat h}}_{{\rm{near - field}}}^{\rm{P}} = {{\bf{0}}_{M_b M_u \times 1}}$.
     \For{$ l_n = 1,2,\cdots,L_{\rm{N}}$}
        \State Update the support set: $ \eta = \mathop {\arg \max }\limits_{n = 1,2, \cdots ,N} \left\| {{\bf{A}}_{\rm{N}}^H\left( {:,n} \right){\bf{r}}} \right\|_2^2   $, $ \Omega_{\rm{N}} = \Omega_{\rm{N}} \bigcup \eta $.
        \State Perform the LMS process: ${\bf{x}} = {\bf{\hat h}}_{{\rm{near - field}}}^{\rm{P}}(\Omega_{\rm{N}})$,
        \For{$ n = 1,2,\cdots,N_b N_u$}
            \State ${\bf{c}} = {\bf{A}}_{\rm{N}}(n,\Omega_{\rm{N}})$, $d = {\bf{y}}(n) $, $ e = d - {\bf{c}} {\bf{x}} $,
            \State $ {\bf{x}} = {\bf{x}} + \mu e {\bf{c}}^H $,
        \EndFor
        \State ${\bf{\hat h}}_{{\rm{near - field}}}^{\rm{P}}(\Omega_{\rm{N}}) = {\bf{x}}$.
        \State Compute the residual vector: $ {\bf{r}} = {\bf{y}} - {\bf{A}}_{\rm{F}}{\bf{\hat h}}_{{\rm{far - field}}}^{\rm{A}} - {\bf{A}}_{\rm{N}}{\bf{\hat h}}_{{\rm{near - field}}}^{\rm{P}} $.
    \EndFor

  \Statex // \textbf{Phase \uppercase\expandafter{\romannumeral3}:} Reconstructing the hybrid-field channel.
    \State Reshape the vectors ${\bf{\hat h}}_{{\rm{far - field}}}^{\rm{A}}$ and ${\bf{\hat h}}_{{\rm{near - field}}}^{\rm{P}}$ to the matrices ${\bf{\hat H}}_{{\rm{far - field}}}^{\rm{A}} \in {\mathbb{C}^{{N_b} \times N_u }}  $ and ${\bf{\hat H}}_{{\rm{near - field}}}^{\rm{P}} \in {\mathbb{C}^{{M_b} \times M_u }}  $.
    \State Reconstruct the hybrid-field channel: ${\bf{\hat H}} = {\bf{F}}_{\rm R}  {{\bf{\hat H}}_{{\rm{far - field}}}^{\rm A}} {\bf{F}}_{\rm T}^H + {{\bf{D}}_{\rm{R}}}{\bf{\hat H}}_{{\rm{near - field}}}^{\rm{P}}{\bf{D}}_{\rm{T}}^H $.

  \end{algorithmic}
\end{algorithm}

\subsection{On-Grid Hybrid-Field SGP Channel Estimation With $ \gamma $}

The specific algorithm of the on-grid hybrid-field SGP channel estimation scheme with $ \gamma $ is succinctly summarized in {\textbf{{Algorithm}} \ref{alg:1}}.
We denote the support sets for the far-field and the near-field path components by $\Omega_{\rm{F}}$ and $\Omega_{\rm{N}}$, respectively, initially set to empty sets.
Let $ {\bf{A}}_{\rm{F}} =  \sqrt \tau  \left( {{{\bf{P}}^T}{\bf{F}}_{\rm{T}}^* \otimes {{\bf{F}}_{\rm{R}}}} \right) $ denote the sensing matrix of the far-field path components.
It is worth noting that employing a near-field sensing matrix in the far-field is feasible.
This is primarily because the far-field channel in \eqref{(5)} is a special case of the near-field channel \eqref{(8)}.
However, due to the increased complexity and substantially higher dimensionality of the near-field sensing matrix compared to its far-field counterpart, we have opted for the far-field sensing matrix in our far-field estimation approach.
The analysis of algorithm's three phases is discussed as follows.

\subsubsection{Phase \uppercase\expandafter{\romannumeral1}}
During this phase, the on-grid hybrid-field SGP iteratively estimates the far-field path components in the angular domain, as shown in Steps $2$-$12$.
Specifically, $ L_{\rm{F}} $ supports, associated with $ L_{\rm{F}} $ far-field path components in the angular domain, are determined over $ L_{\rm{F}} $ iterations.
For the iteration $ l_f \in [1,2,\cdots,L_{\rm{F}} ]$, the $ l_f $-th far-field support $\eta$ is acquired
by calculating the inner products between the sensing matrix $ {\bf{A}}_{\rm{F}} $ and the residual vector $ {\bf{r}} $, and subsequently identifying the most correlative column index of  $ {\bf{A}}_{\rm{F}} $ with $ {\bf{r}} $, as shown in Step $3$.
Following this, the far-field support set will be updated in Step $3$.
Leveraging the far-field support set $ \Omega_{\rm{F}} $, the LMS algorithm is employed to estimate the far-field sparse vector $ {\bf{\hat h}}_{{\rm{far - field}}}^{\rm{A}} $ for the current iteration with step size $ \mu $.
Then, we update the residual vector $ {\bf{r}} $ by subtracting the contributions of the estimated far-field path components, as shown in Step $11$.
Subseguently, after $ L_{\rm{F}} $ iterations, we can finally obtain the estimated far-field path components $ {\bf{\hat h}}_{{\rm{far - field}}}^{\rm{A}} $.

\subsubsection{Phase \uppercase\expandafter{\romannumeral2}}
In this phase, the near-field sensing matrix $ {\bf{A}}_{\rm{N}} =  \sqrt \tau  \left( {{{\bf{P}}^T}{\bf{D}}_{\rm{T}}^* \otimes {{\bf{D}}_{\rm{R}}}} \right) $ is exploited to estimate the near-field path components in the polar domain.
There are two main differences between this phase (Steps $14$-$24$) and the first phase (Steps $2$-$12$).
The first distinction is that the sensing matrix employed in this phase is the near-field sensing matrix $ {\bf{A}}_{\rm{N}} =  \sqrt \tau  \left( {{{\bf{P}}^T}{\bf{D}}_{\rm{T}}^* \otimes {{\bf{D}}_{\rm{R}}}} \right) $, while the sensing matrix employed in the first phase is the far-field sensing matrix $ {\bf{A}}_{\rm{F}} =  \sqrt \tau  \left( {{{\bf{P}}^T}{\bf{F}}_{\rm{T}}^* \otimes {{\bf{F}}_{\rm{R}}}} \right) $.
The second difference lies in the residual update at this stage, which requires eliminating the contributions of both the estimated near-field paths components and the estimated far-field paths components from the residual.

\subsubsection{Phase \uppercase\expandafter{\romannumeral3}}
After estimating the vectors ${\bf{\hat h}}_{{\rm{far - field}}}^{\rm{A}}$ and ${\bf{\hat h}}_{{\rm{near - field}}}^{\rm{P}}$, they are reshaped into matrices ${\bf{\hat H}}_{{\rm{far - field}}}^{\rm{A}}$ and ${\bf{\hat H}}_{{\rm{near - field}}}^{\rm{P}}$, respectively, constructing the entire estimated channel $ {\bf{\hat H}} $ based on \eqref{(12)}, as shown in Step $26$.
It is worth noting that the key difference between the proposed on-grid hybrid-field SGP and the hybrid-field OMP in \cite{[17]} lies in their approach to estimate the sparse vector in the each iteration.
Specifically, the proposed hybrid-field SGP exploits the LMS algorithm, differing from the LS algorithm employed in the hybrid-field OMP.
In this way, the on-grid hybrid-field SGP is expected to outperform the hybrid-field OMP which will be verified in Section \ref{Simulation Results} via simulations. 

Note that the above hybrid-field SGP heavily relies on the prior known information of $ \gamma $.
Yet, in practice, the specific number of the channel paths in far-field and near-field may not be always available such that the value of $ \gamma $ is unknown.
Therefore, in this scenario, the hybrid-field OMP in \cite{[17]} and the proposed on-grid hybrid-field SGP cannot be directly applied.
Moreover, the algorithms previously mentioned, including the hybrid-field OMP in \cite{[17]}, the SD-OMP in \cite{[18]}, the practical OMP in \cite{[19]}, and the proposed on-grid hybrid-field SGP, all operate under the assumption that the distances and angles correspond precisely to sampled points in the polar and angular domains, respectively.
This assumption inherently limits the estimation accuracy of these algorithms to the density of grid sampling points.
Building upon the insights from \cite{[4]} and \cite{[19]}, we will introduce an off-grid hybrid-field SGP channel estimation scheme in the following section.
This new scheme does not require the prior known information of $ \gamma $ and offers enhanced capabilities for refining channel parameters.

\subsection{Off-Grid Hybrid-Field SGP Channel Estimation Without $ \gamma $}

For facilitating the design of the off-grid hybrid-field SGP without $ \gamma $, we first reformulate the vector $ {{\bf{y}}_t} $ as
\begin{equation}\label{(15)}
\begin{aligned}
{{\bf{y}}_t} &= {\bf{A}}_{\rm{F}}{\bf{h}}_{{\rm{far - field}}}^{\rm{A}} + {\bf{A}}_{\rm{N}}{\bf{h}}_{{\rm{near - field}}}^{\rm{P}} + {{\bf{n}}_t}\\
&={\bf{A}}{\bf{\bar{h}}}  + {{\bf{n}}_t},
\end{aligned}
\end{equation}
where $ {\bf{A}}_{\rm{F}} =  \sqrt \tau  \left( {{{\bf{P}}^T}{\bf{F}}_{\rm{T}}^* \otimes {{\bf{F}}_{\rm{R}}}} \right)\in {\mathbb{C}^{{\left(N_b N_u\right)} \times \left(N_b N_u\right)}}  $, $ {\bf{A}}_{\rm{N}} =  \sqrt \tau  \left( {{{\bf{P}}^T}{\bf{D}}_{\rm{T}}^* \otimes {{\bf{D}}_{\rm{R}}}} \right)\in {\mathbb{C}^{{\left(N_b N_u\right)} \times {\left(M_b M_u\right)}}}  $, $ {\bf{A}} = \left[  {\bf{A}}_{\rm{F}},  {\bf{A}}_{\rm{N}}  \right] \in {\mathbb{C}^{{N_bN_u} \times {\left(N_b N_u+M_b M_u\right)} }} $, and $ {\bf{\bar{h}}} = \left[  {\bf{h}}_{{\rm{far - field}}}^{\rm{A}},  {\bf{h}}_{{\rm{near - field}}}^{\rm{P}}  \right]\in {\mathbb{C}^{{\left(N_b N_u+M_b M_u\right)} \times 1}} $.
Then, the procedures of the off-grid hybrid-field SGP is presented in {\textbf{{Algorithm}} \ref{alg:11}}.
Specifically, the off-grid hybrid-field SGP can also be divided into three phases.

\begin{algorithm}[t]
  \fontsize{11}{13}\selectfont
  \caption{The proposed Off-grid Hybrid-field SGP Channel Estimation Scheme without Prior Knowledge $\gamma$. }
  \label{alg:11}
  \begin{algorithmic}[1]
  \Require ${\bf{y}}_t$, ${\bf{A}}_{\rm{F}}$, ${\bf{A}}_{\rm{N}}$, ${\bf{A}}$, $L$, $\nu$
  \Ensure ${\bf{\hat H}} $
  {\Statex  \textbf{Initialization:} $ \gamma=1 $, $ \Omega_{\rm{F}}=\emptyset $, $ {\bf{r}} = {\bf{y}} $}.
  \Statex // \textbf{Phase \uppercase\expandafter{\romannumeral1}:} The coarse estimation of the far-field path components in the angular domain.

    \State $ {\bf{\hat h}}_{{\rm{far - field}}}^{\rm{A}} = {{\bf{0}}_{N_b N_u \times 1}}$.
    \For{$ l = 1,2,\cdots,L$}
        \State Store residual vector: ${\bf{R}}(:,l) = {\bf{r}}$.
        \State $ \eta = \mathop {\arg \max }\limits_{n = 1,2, \cdots ,N_b N_u} \left\| { {\bf{A}}_{\rm{F}}^H\left( {:,n} \right){\bf{r}}} \right\|^2   $, $ \Omega_{\rm{F}} = \Omega_{\rm{F}} \bigcup \eta $.
        \State Perform the LMS process: ${\bf{x}} = {\bf{\hat h}}_{{\rm{far - field}}}^{\rm{A}}(\Omega_{\rm{F}})$,
        \For{$ n = 1,2,\cdots,N_b N_u$}
            \State ${\bf{c}} = {\bf{A}}_{\rm{F}}(n,\Omega_{\rm{F}})$, $ e = {\bf{y}}_t(n) - {\bf{c}} {\bf{x}} $, $ {\bf{x}} = {\bf{x}} + \nu e {\bf{c}}^H $,
        \EndFor
        \State ${\bf{\hat h}}_{{\rm{far - field}}}^{\rm{A}}(\Omega_{\rm{F}}) = {\bf{x}}$.
        \State Compute the residual vector: $ {\bf{r}} = {\bf{y}} - {\bf{A}}_{\rm{F}} {\bf{\hat h}}_{{\rm{far - field}}}^{\rm{A}} $.
    \EndFor

  \Statex // \textbf{Phase \uppercase\expandafter{\romannumeral2}:} The precise estimation of the near-field path components in the polar domain.
    \State $ {\bf{r}}_m =  {\bf{r}} $, $ {\bf{\widetilde h}} = {\bf{\hat h}}_m = {{\bf{0}}_{(N_b N_u+M_b M_u) \times 1}} $, $ {\bf{ \widetilde{h}}} (1:N_b N_u)= {\bf{\hat h}}_{{\rm{far - field}}}^{\rm{A}} $, $ \Omega = \emptyset $.
     \For{$ \gamma = \frac{{L - 1}}{L}: - \frac{1}{L}:0$}
         \State $ \Omega = \Omega_{\rm{F}}(1:\gamma L) $, $ {\bf{r}} =  {\bf{R}}(:,\gamma L +1) $.
         \For{$ l = 1 : (1-\gamma)L $ }
            \State $ \eta = \mathop {\arg \max }\limits_{n = 1,2, \cdots ,M_b M_u} \left\| {{\bf{A}}_{\rm{N}}^H\left( {:,n} \right){\bf{r}}} \right\|^2   $, $ \Omega = \Omega \bigcup \eta $.
            \State Perform the LMS process: ${\bf{x}} = {\bf{ \widetilde{h}}}(\Omega)$,
            \For{$ n = 1,2,\cdots,N_b N_u$}
                \State ${\bf{c}} = {\bf{A}}(n,\Omega)$, $ e = {\bf{y}}_t(n) - {\bf{c}} {\bf{x}} $, $ {\bf{x}} = {\bf{x}} + \nu e {\bf{c}}^H $,
            \EndFor
            \State ${\bf{ \widetilde{h}}}(\Omega) = {\bf{x}}$.
            \State Compute the residual vector: $ {\bf{r}} = {\bf{y}} -  {\bf{A}} {\bf{ \widetilde{h}}}$.
         \EndFor
         \If{$ \left\| {\bf{r}} \right\| \le \left\| {{{\bf{r}}_m}} \right\| $}
            \State Update the minimal residual vector: $ {\bf{r}}_m =  {\bf{r}} $.
            \State Update the sparse channel vector: $ {\bf{\hat h}}_m = {\bf{ \widetilde{h}}}$.
         \EndIf
    \EndFor
  \Statex // \textbf{Phase \uppercase\expandafter{\romannumeral3}:} {{Refining channel parameters.}}
    \For{$ n = 1, 2, \cdots, N_{\rm iter}$}
        \State {{Update the angles by ${{{\bm{\hat \theta }}}^n} = {{{\bm{\hat \theta }}}^{n - 1}} - {\bm{\zeta}}_1^T{\nabla _{{\bm{\hat \theta }}}}{\mathcal L}({{{\bf{\hat r}}}^{n - 1}},{\bm{\hat \theta }}){|_{{\bm{\hat \theta }} = {{{\bm{\hat \theta }}}^{n - 1}}}}$ as in \eqref{(1004)}.}}
        \State {{Update the distances by $ \frac{1}{{{{{\bf{\hat r}}}^n}}} = \frac{1}{{{{{\bf{\hat r}}}^{n - 1}}}} - {\bm{\zeta}}_2^T{\nabla _{\frac{1}{{{\bf{\hat r}}}}}}{\mathcal L}({\bf{\hat r}},{{{\bm{\hat \theta }}}^n}){|_{{\bf{\hat r}} = {{{\bf{\hat r}}}^{n - 1}}}} $ as in \eqref{(1005)}.}}
        \State {{Update the path gains ${\bf{\hat g}}^{n}$ as in \eqref{(1002)}.}}
    \EndFor
    \State {{Update the sparse channel vector: ${\bf{\hat h}} = {\bf{\hat A}}^{N_{\rm iter}}  {\bf{\hat g}}^{N_{\rm iter}}$.}}
    \State {{Reshape the vectors $ {\bf{\hat h}}  $ to the channel matrix $ {\bf{\hat H}} $.}}


  \end{algorithmic}
\end{algorithm}

\subsubsection{Phase \uppercase\expandafter{\romannumeral1}}
In this phase, we assume $ \gamma=1 $ to perform the coarse estimation in far-field, assuming that all the path components follow the far-field model.
Consequently, the SGP algorithm can be employed to perform the channel estimation, as shown in Steps $2$-$11$.
For the $l$-th iteration of {\textbf{{Algorithm}} \ref{alg:11}}, in Steps $3$, the residual vectors from the previous iteration are stored at $ {\bf{R}} \in {\mathbb{C}^{{\left(N_b N_u\right)} \times 1}}  $ with $ {\bf{R}}(:,l) = {\bf{r}} $, aiding in the adjustment of $ \gamma $ in the second phase.
Next, the $l$-th far-field support is detected via the orthogonal pursuing process, which is similar to {\textbf{{Algorithm}} \ref{alg:1}}, as shown in Step $4$.
Based on the updated far-field support set, the sparse vector ${\bf{\hat h}}_{{\rm{far - field}}}^{\rm{A}}(\Omega_{\rm{F}})$ of the current iteration can be obtained by the LMS algorithm with the step size $ \nu $, as shown in Steps $5$-$9$.
Subsequently, the residual vector of the $l$-th iteration is updated in Step $10$.

\subsubsection{Phase \uppercase\expandafter{\romannumeral2}}
During this phase, we adjust $\gamma$ to perform precise estimation in the near-field.
We denote the minimal of the residual vector by $ {\bf{r}}_m  $ with the initialization of $ {\bf{r}}_m={\bf{r}} $ and the sparse channel vector corresponding to $ {\bf{r}}_m $ by $ {\bf{\hat h}}_m $.
Let $ {\bf{\widetilde h}} $ denote the estimation of the channel vector $ {\bf{\bar{h}}} $, which is initialized as $ {\bf{\widetilde h}} = {{\bf{0}}_{(N_b N_u+M_b M_u) \times 1}} $, $ {\bf{ \widetilde{h}}} (1:N_b N_u)= {\bf{\hat h}}_{{\rm{far - field}}}^{\rm{A}} $.
Let $ \Omega $ denote the set which includes the combination of the far-field and near-field supports.
{{Then, in Step $13$, $ \gamma $ is adjusted from $1$ to $0$ with a step size of $- {1 \mathord{\left/ {\vphantom {1 L}} \right. \kern-\nulldelimiterspace} L}$ to traverse all possibilities, aiming to acquire the minimal residual vector.
It is worth noting that the initial value of $\gamma$ does not influence the results of the algorithm, as it iteratively explores all the values of $\gamma$.}}
Moreover, for an arbitrary value of $\gamma$, the corresponding $ \gamma L $ far-field supports and residual
vector have been stored in $ \Omega_{\rm{F}}(1:\gamma L) $ and $ {\bf{R}}(:,\gamma L +1) $ in the first phase, respectively.
Therefore, the remaining task is to search for the $ (1-\gamma)L $ near-field supports in the polar domain.
Specifically, we first initialize $ \Omega $ and $ {\bf{r}} $ as $ \Omega = \Omega_{\rm{F}}(1:\gamma L) $ and $ {\bf{r}} =  {\bf{R}}(:,\gamma L +1) $ in Step $14$, respectively.
Next, the SGP algorithm is exploited to perform the remaining channel estimation, as shown in Steps $15$-$23$.
Notable, the goal of Step $16$ is to find the near-field supports and update them to the support set $ \Omega $.
In addition, Steps $17$-$21$ aim to recover the sparse vector $ {\bf{\bar{h}}} $ from the signal $ {{\bf{y}}_t} ={\bf{A}}{\bf{\bar{h}}}  + {{\bf{n}}_t} $ following the CS framework.
Then, we update the minimal residual vector $ {\bf{r}}_m  $ and the sparse channel vector $ {\bf{\hat h}}_m $ by comparing the norms of $ {\bf{r}}_m $ and $ {\bf{r}} $.

\begin{table*}[t]
  \begin{center}
  \fontsize{7}{13}\selectfont
  \renewcommand{\arraystretch}{2}

    \caption{Computational Complexity for Different Channel Estimation Schemes Based on the Number of complex-Valued Multiplications\protect\footnotemark[5].}

      \label{table}
 \begin{threeparttable}
    \begin{tabular}  {
    !{\vrule width1.2pt}  c
    !{\vrule width1.2pt}  c
    !{\vrule width1.2pt}  c
    !{\vrule width1.2pt}  c
    !{\vrule width1.2pt}  c
    !{\vrule width1.2pt}} 

      \Xhline{1.2pt}
         \bf Scheme
        &  \makecell{ \bf The update of  \\ \bf the support set  }
        &  { \makecell{ \bf The LS or LMS process } }
        & \makecell{ \bf The update of \\  \bf the residual  }
        & \makecell{ \bf The reconstruction \\ \bf of the channel  }  \\
      \Xhline{1.2pt}

      \makecell{The hybrid-field \\ OMP with $\gamma$ \cite{[17]}}
      &  \makecell{   $ \frac{L}{2}{N_b}{N_u}( {{N_b}{N_u}}$\\$ + {M_b}{M_u} )$    }
      &  \makecell{ $ 2 \sum\limits_{l = 1}^{ {L \mathord{\left/ {\vphantom {L 2}} \right. \kern-\nulldelimiterspace} 2}} { ( {\frac{1}{2}{l^3} + \frac{3}{2}{l^2}}}$\\ $+ {{  2{N_b}{N_u}{l^2} + {N_b}{N_u}l} )}  $ }
      & \makecell{ $ \frac{L}{2}{N_b}{N_u}( {{N_b}{N_u}}$\\$ + {M_b}{M_u} )$  }
      & \makecell{ $ {N_b}{N_u}\left( {{N_b} + {N_u}} \right)$ \\$ + {N_b}{M_u}\left( {{N_u} + {M_b}} \right)$   } \\
      \hline

      \makecell{The on-grid \\ hybrid-field SGP}
      & \makecell{   $ \frac{L}{2}{N_b}{N_u}( {{N_b}{N_u}}$\\$ + {M_b}{M_u} )$    }
      &  \makecell{ $2\sum\limits_{l = 1}^{{L \mathord{\left/{\vphantom {L 2}} \right.\kern-\nulldelimiterspace} 2}} {\left[ {{N_b}{N_u}\left( {2l + 1} \right)} \right]} $  }
      & \makecell{   $ \frac{L}{2}{N_b}{N_u}( {{N_b}{N_u}}$\\$ + {M_b}{M_u} )$    }
      & \makecell{ $ {N_b}{N_u}\left( {{N_b} + {N_u}} \right)$ \\$ + {N_b}{M_u}\left( {{N_u} + {M_b}} \right)$   }\\
      \hline

      \makecell{The hybrid-field\\ OMP without $\gamma$ \cite{[19]} }
      & \makecell{ $LN_b^2N_u^2$\\$+ \frac{1}{2}L\left( {1 + L}  \right) $\\$\times{N_b}{N_u}{M_b}{M_u}$ }
      &  \makecell{ $\sum\limits_{l = 1}^L {( {\frac{1}{2}{l^3} + \frac{3}{2}{l^2} }}  $  $+{{ 2{N_b}{N_u}{l^2} + {N_b}{N_u}l} )}$ \\ $+ \sum\limits_{i = 0}^{L - 1} {\sum\limits_{l = 1}^{L - i} {( {\frac{1}{2}{l^3} + \frac{3}{2}{l^2} }}}$ $+{{{{ 2{N_b}{N_u}{l^2} + {N_b}{N_u}l} )} }} $  }
      & \makecell{ $LN_b^2N_u^2$\\$+ \frac{L}{2}\left( {1 + L} \right)[ {{N_b}{N_u} }$ \\ ${\times  ( {{N_b}{N_u} + {M_b}{M_u}} )} ]$  }
      & \makecell{ $ {N_b}{N_u}\left( {{N_b} + {N_u}} \right)$ \\$ + {N_b}{M_u}\left( {{N_u} + {M_b}} \right)$   } \\
      \hline

      \makecell{The off-grid \\ hybrid-field SGP }
      & \makecell{ $LN_b^2N_u^2$\\$+ \frac{1}{2}L\left( {1 + L}  \right) $\\$\times{N_b}{N_u}{M_b}{M_u}$ }
      & \makecell{$\sum\limits_{l = 1}^L {\left[ {{N_b}{N_u}\left( {2l + 1} \right)} \right]}  $\\$+\sum\limits_{i = 0}^{L - 1} {\sum\limits_{l = 1}^{L - i} {\left[ {{N_b}{N_u}\left( {2l + 1} \right)} \right]} } $  }
      &\makecell{ $LN_b^2N_u^2$\\$+ \frac{L}{2}\left( {1 + L} \right)[ {{N_b}{N_u} }$ \\ $\times{  ( {{N_b}{N_u} + {M_b}{M_u}} )} ]$  }
      & \makecell{ $  4{N_{\rm iter}}(N_b^2N_u^2   $ \\   $  + 7{L^2}{N_b}{N_u} + 8L{N_b}{N_u})  $   }\\
      \Xhline{1.2pt}

    \end{tabular}
   \begin{tablenotes}[para,flushleft]
         \item In this table, $ N_b $ represents the number of antennas for the BS, with the corresponding number of sampled grids denoted as $M_b$. Similarly, $ N_u $ is the number of antennas for the UE, with the corresponding number of sampled grids denoted as $M_u$. The parameter $L$ signifies the total number of channel paths. Additionally, it is important to note that the computational complexity of all the schemes in the table is calculated under the scenario where $\gamma = 0.5$. This implies that the number of far-field path components, denoted as $L_{\rm{F}}$, is equal to the number of near-field path components, denoted as $L_{\rm{N}}$.
   \end{tablenotes}
  \end{threeparttable}
  \end{center}
\end{table*}


\subsubsection{Phase \uppercase\expandafter{\romannumeral3}}

According to the sensing matrices and the support set $ \Omega $, we can obtain the initial value of the estimated distances ${{{\bf{\hat r}}}_{{\rm{R}},{\rm{near}}}} = [{{\hat r}_{{\rm{R}},1}},{{\hat r}_{{\rm{R}},2}}, \cdots ,{{\hat r}_{{\rm{R}},{L_{\rm{N}}}}}]^T$, ${{{\bf{\hat r}}}_{{\rm{T}},{\rm{near}}}} = [{{\hat r}_{{\rm{T}},1}},{{\hat r}_{{\rm{T}},2}}, \cdots ,{{\hat r}_{{\rm{T}},{L_{\rm{N}}}}}]^T$,
angles ${{{\bm{\hat \theta }}}_{{\rm{R}},{\rm{near}}}} = [{{\hat \theta }_{{\rm{R}},1}},{{\hat \theta }_{{\rm{R}},2}}, \cdots ,{{\hat \theta }_{{\rm{R}},{L_{\rm{N}}}}}]^T$, ${{{\bm{\hat \theta }}}_{{\rm{T}},{\rm{near}}}} = [{{\hat \theta }_{{\rm{T}},1}},{{\hat \theta }_{{\rm{T}},2}}, \cdots ,{{\hat \theta }_{{\rm{T}},{L_{\rm{N}}}}}]^T$, ${{{\bm{\hat \theta }}}_{{\rm{R}},{\rm{far}}}} = [{{\hat \theta }_{{\rm{R}},1}},{{\hat \theta }_{{\rm{R}},2}}, \cdots ,{{\hat \theta }_{{\rm{R}},{L_{\rm{F}}}}}]^T$, ${{{\bm{\hat \theta }}}_{{\rm{T}},{\rm{far}}}} = [{{\hat \theta }_{{\rm{T}},1}},{{\hat \theta }_{{\rm{T}},2}}, \cdots ,{{\hat \theta }_{{\rm{T}},{L_{\rm{F}}}}}]^T$
and complex path gains ${\bf{\hat g}} = {{{\bf{\hat h}}}_m}(\Omega) \in {\mathbb{C}^{{L} \times 1}} $.
Then, these estimated channel parameters can be utilized to construct the matrix
\begin{equation}
\begin{aligned}\label{(990)}
\!{{{\bf{\hat A}}}_{\rm{F}}} =& \sqrt \tau  \sqrt {\frac{p}{{{N_u}}}} {\big[} {{\bf{a}}^*}({{\hat \theta }_{{\rm{T}},{\rm{1}}}}) \otimes {\bf{a}}({{\hat \theta }_{{\rm{R}},{\rm{1}}}}),\\
&{{\bf{a}}^*}({{\hat \theta }_{{\rm{T}},2}}) \otimes {\bf{a}}({{\hat \theta }_{{\rm{R}},2}}), \cdots ,{{\bf{a}}^*}({{\hat \theta }_{{\rm{T}},{L_{\rm{F}}}}}) \otimes {\bf{a}}({{\hat \theta }_{{\rm{R}},{L_{\rm{F}}}}}) {\big]},
\end{aligned}
\end{equation}
\begin{equation}
\begin{aligned}\label{(991)}
\!\!\!\!\!{{{\bf{\hat A}}}_{\rm{N}}} =& \sqrt \tau  \sqrt {\frac{p}{{{N_u}}}} {\big[}{{\bf{a}}^*}({{\hat r}_{{\rm{T}},1}},{{\hat \theta }_{{\rm{T}},{\rm{1}}}}) \otimes {\bf{a}}({{\hat r}_{{\rm{R}},1}},{{\hat \theta }_{{\rm{R}},{\rm{1}}}}),\\
&{{\bf{a}}^*}({{\hat r}_{{\rm{T}},2}},{{\hat \theta }_{{\rm{T}},2}}) \otimes {\bf{a}}({{\hat r}_{{\rm{R}},2}},{{\hat \theta }_{{\rm{R}},2}}), \cdots ,\\
&{{\bf{a}}^*}({{\hat r}_{{\rm{T}},{L_{\rm{N}}}}},{{\hat \theta }_{{\rm{T}},{L_{\rm{N}}}}}) \otimes {\bf{a}}({{\hat r}_{{\rm{R}},{L_{\rm{N}}}}},{{\hat \theta }_{{\rm{R}},{L_{\rm{N}}}}}){\big]},
\end{aligned}
\end{equation}
where
$ {{{\bf{\hat A}}}_{\rm{F}}} \in {\mathbb{C}^{ {{N_b}{N_u}} \times {L_{\rm F}}  }}  $,
${{{\bf{\hat A}}}_{\rm{N}}} \in {\mathbb{C}^{ {{N_b}{N_u}} \times {L_{\rm N}}  }} $, and $ {\bf{\hat A}} = [{{{\bf{\hat A}}}_{\rm{F}}},{{{\bf{\hat A}}}_{\rm{N}}}]  \in {\mathbb{C}^{ {{N_b}{N_u}} \times {L}  }}  $.
Therefore, the recovered channel vector is ${\bf{\hat h}} = {\bf{\hat A\hat g}}$.
To maximize the likelihood, we alternatively optimize the angles, distances, and path gains, which is formulated as
\begin{equation}
\begin{aligned}\label{(1001)}
\mathop {\min }\limits_{{\bf{\hat g}},{{{\bf{\hat r}}}_{{\rm{R}},{\rm{near}}}},{{{\bf{\hat r}}}_{{\rm{T}},{\rm{near}}}},{{{\bm{\hat \theta }}}_{{\rm{R}},{\rm{near}}}},{{{\bm{\hat \theta }}}_{{\rm{T}},{\rm{near}}}},{{{\bm{\hat \theta }}}_{{\rm{R}},{\rm{far}}}},{{{\bm{\hat \theta }}}_{{\rm{T}},{\rm{far}}}}} {\left\| {{{\bf{y}}_t} - {\bf{\hat A\hat g}}} \right\|^2}.
\end{aligned}
\end{equation}

Since the optimization problem \eqref{(1001)} is non-convex, the alternating minimization method is utilized to obtain an effective solution.
To express conciseness, let $ {{\bf{\hat r}}} = [{{{\bf{\hat r}}}_{{\rm{R}},{\rm{near}}}}^T, {{{\bf{\hat r}}}_{{\rm{T}},{\rm{near}}}}^T]^T $, $ {{\bm{\hat \theta }}} = [ {{{\bm{\hat \theta }}}_{{\rm{R}},{\rm{near}}}}^T, {{{\bm{\hat \theta }}}_{{\rm{T}},{\rm{near}}}}^T, {{{\bm{\hat \theta }}}_{{\rm{R}},{\rm{far}}}}^T, {{{\bm{\hat \theta }}}_{{\rm{T}},{\rm{far}}}}^T  ]^T $.
For fixed ${{\bf{\hat r}}}$ and ${{\bm{\hat \theta }}}$, the optimal solution for $ {\bf{\hat g}} $ is given by
\begin{equation}
\begin{aligned}\label{(1002)}
{\bf{\hat g}}^{\rm opt} = {\bf{\hat A }}^{\dagger}{{\bf{y}}_t}.
\end{aligned}
\end{equation}

Plug \eqref{(1002)} into \eqref{(1001)}, the maximum-likelihood problem can be reformulated as
\begin{equation}
\begin{aligned}\label{(1003)}
&\mathop {\min }\limits_{{\bf{\hat g}},{\bf{\hat r}},{\bm{\hat \theta }}} {\left\| {{{\bf{y}}_t} - {\bf{\hat A\hat g}}} \right\|^2}
 \Leftrightarrow \mathop {\min }\limits_{{\bf{\hat r}},{\bm{\hat \theta }}} {\rm{tr}}\left\{ {{\bf{y}}_t^H{{\left( {{\bf{1}} - {\bf{\Psi }} } \right)}^H}\left( {{\bf{1}} - {\bm{\Psi }} } \right){{\bf{y}}_t}} \right\}\\
 &\mathop  \Leftrightarrow \limits^{(a)} \mathop {\min }\limits_{{\bf{\hat r}},{\bm{\hat \theta }}} {\mathcal L}({{\bf{\hat r}}}, {{\bm{\hat \theta }}} ) =  - {\rm{tr}}\left\{ {{\bf{y}}_t^H {\bm{\Psi }} {{\bf{y}}_t}} \right\},
\end{aligned}
\end{equation}
where $ {\bf{\Psi }} = {\bf{\hat A}}{{{\bf{\hat A}}}^\dag } $, and $(a)$ holds by $ {{\bf{\Psi }}^H}{\bf{\Psi }} = {\bf{\Psi }} $.
Then, we can utilize an iterative gradient descent to optimize the new minimization problem.
In the $n$-th iteration, the angles are updated as
\begin{equation}
\begin{aligned}\label{(1004)}
{{{\bm{\hat \theta }}}^n} = {{{\bm{\hat \theta }}}^{n - 1}} - {\bm{\zeta}}_1^T{\nabla _{{\bm{\hat \theta }}}}{\mathcal L}({{{\bf{\hat r}}}^{n - 1}},{\bm{\hat \theta }}){|_{{\bm{\hat \theta }} = {{{\bm{\hat \theta }}}^{n - 1}}}},
\end{aligned}
\end{equation}
where $ {\bm{\zeta}}_1 $ is the step length vector.
Note that for different angles $ {{{\bm{\hat \theta }}}_{{\rm{R}},{\rm{near}}}}$, ${{{\bm{\hat \theta }}}_{{\rm{T}},{\rm{near}}}}$, ${{{\bm{\hat \theta }}}_{{\rm{R}},{\rm{far}}}}$, and ${{{\bm{\hat \theta }}}_{{\rm{T}},{\rm{far}}}}  $, the update strategy is the same, but the gradient and step length are different.
As for the near-field distances, the authors of \cite{[4]} had demonstrated that $ 1/r $ is uniformly sampled in the polar domain.
In this way, we define $\frac{1}{{{\bf{\hat r}}}} = [\frac{1}{{{{\hat r}_{{\rm{R}},1}}}},\frac{1}{{{{\hat r}_{{\rm{R}},2}}}}, \cdots ,\frac{1}{{{{\hat r}_{{\rm{R}},{L_{\rm{N}}}}}}},\frac{1}{{{{\hat r}_{{\rm{T}},1}}}},\frac{1}{{{{\hat r}_{{\rm{T}},2}}}}, \cdots ,\frac{1}{{{{\hat r}_{{\rm{T}},{L_{\rm{N}}}}}}}]^T$, which are updated as
\begin{equation}
\begin{aligned}\label{(1005)}
\frac{1}{{{{{\bf{\hat r}}}^n}}} = \frac{1}{{{{{\bf{\hat r}}}^{n - 1}}}} - {\bm{\zeta}}_2^T{\nabla _{\frac{1}{{{\bf{\hat r}}}}}}{\mathcal L}({\bf{\hat r}},{{{\bm{\hat \theta }}}^n}){|_{{\bf{\hat r}} = {{{\bf{\hat r}}}^{n - 1}}}},
\end{aligned}
\end{equation}
where $ {\bm{\zeta}}_2 $ is the step length vector for the inverse of near-field distances.
The step length vector $ {\bm{\zeta}}_1 $ and $ {\bm{\zeta}}_2 $ are chosen by Armijo backtracking line search \cite{[4]} to guarantee the objective function is non-increasing.
The gradient $ \nabla {\mathcal L}({\bf{\hat r}},{{{\bm{\hat \theta }}}})$ is given in the appendix.
Finally, the refined channel parameters can be utilized to reconstruct the estimation of the channel, as shown in Steps $30$-$35$ in {\bf Algorithm 2}.

\subsection{Computational Complexity}

In this subsection, we analyze the number of complex-valued multiplications of the proposed channel estimation schemes.
To present the differences in computational complexity between the hybrid-field SGP and hybrid-field OMP more intuitively, we assume $ {L_{\rm{F}}} = {L_{\rm{N}}} = {L \mathord{\left/ {\vphantom {L 2}} \right. \kern-\nulldelimiterspace} 2} $ for the complexity analysis.
We then divide the hybrid-field channel estimation schemes into four parts: the update of the support set, the LS or LMS process, the update of the residual, and the reconstruction of the channel.

For the on-grif hybrid-field SGP with $\gamma$ in {\textbf{{Algorithm}} \ref{alg:1}}, the update of the support set is in Step $3$ and Step $15$ and its number of complex-valued multiplications is
$    \frac{L}{2}{N_b}{N_u}\left( {{N_b}{N_u} + {M_b}{M_u}} \right). $
The LMS process is in Steps $4$-$10$ and Steps $16$-$21$ and its number of complex-valued multiplications is
$ 2\sum\nolimits_{l = 1}^{{L \mathord{\left/
 {\vphantom {L 2}} \right.  \kern-\nulldelimiterspace} 2}} {\left[ {{N_b}{N_u}\left( {2l + 1} \right)} \right]}. $
The update of the residual is in Step $11$ and Step $23$ and its number of complex-valued multiplications is
$  \frac{L}{2}{N_b}{N_u}\left( {{N_b}{N_u} + {M_b}{M_u}} \right). $
Finally, the reconstruction of the channel is in Steps $25$-$26$ and its number of complex-valued multiplications is
$ {N_b}{N_u}\left( {{N_b} + {N_u}} \right) + {N_b}{M_u}\left( {{N_u} + {M_b}} \right). $

As for the off-grid hybrid-field SGP without $\gamma$ in {\textbf{{Algorithm}} \ref{alg:11}}, the update of the support set is in Step $4$ and Step $16$ and its number of complex-valued multiplications is
$ LN_b^2N_u^2 + \frac{1}{2}L\left( {1 + L} \right){N_b}{N_u}{M_b}{M_u}. $
The LMS process is in Steps $5$-$9$ and Steps $17$-$21$ and its number of complex-valued multiplications is
$ \sum_{l = 1}^L {\left( {{N_b}{N_u}\left( {2l + 1} \right)} \right)}  + \sum_{\gamma  = 0}^{L - 1} {\sum_{l = 1}^{\left( {L - \gamma } \right)} {\left( {{N_b}{N_u}\left( {2l + 1} \right)} \right)} }. $
The update of the residual is in Step $10$ and Step $22$ and its number of complex-valued multiplications is
$ LN_b^2N_u^2 + \frac{L}{2}\left( {1 + L} \right)\left[ {{N_b}{N_u}\left( {{N_b}{N_u} + {M_b}{M_u}} \right)} \right]. $
Finally, the reconstruction of the channel is in Steps $30$-$35$ and its number of complex-valued multiplications also is
$ 4{N_{\rm iter}}\left( {N_b^2N_u^2 + 7{L^2}{N_b}{N_u} + 8L{N_b}{N_u}} \right). $

Therefore, the complexity of all the considered channel estimation schemes are summarized in TABLE \ref{table}.
From TABLE \ref{table}, we can observe that the proposed on-grid hybrid-field SGP achieves a lower computational complexity compared to the hybrid-field OMP with $\gamma$ due to the employment of the LMS algorithm.
Similarly, the off-grid hybrid-field SGP also demonstrates reduced computational complexity in comparison to the hybrid-field OMP without $\gamma$.

It is important to highlight that the proposed schemes are also applicable to XL-MIMO systems employing uniform planar array (UPA)-based BS.
When adapting these schemes to such systems, three key areas require careful consideration.
Firstly, by adopting Cartesian coordinates as a reference, channel modeling will shift from a two-dimensional framework, characterized by the $x$ and $y$ axes, to a more intricate three-dimensional model that also incorporates the $z$-axis.
Moreover, the design of codebooks for both far-field and near-field scenarios necessitates reconsideration and customization in light of the new channel model.
This adjustment is imperative to ensure that the codebooks can accurately represent all possible spatial channels.
Lastly, the adoption of the alternating minimization method in this context will result in an escalation in the number of parameters that need refining, reflecting the increased intricacy of the model.
Despite these challenges, the foundational principles of the proposed algorithms remain applicable, enabling effective channel estimation to be conducted within this more complex framework.

\section{Simulation Results}\label{Simulation Results}

In this section, we compare the performance of the proposed channel estimation schemes and the exiting schemes.
The simulation parameters are provided as follows: the carrier wavelength is $ \lambda = 0.01 $ meters, corresponding to a carrier frequency of $ f = 30  $ GHz.
We assume that hybrid-field channel in \eqref{3} consists of $ L = 10 $ path components.
The pilot length is set to $ \tau = 1 $ for simplicity.
We set $ \kappa = 10 $, indicating that the gain of the LoS path and the NLoS paths are generated as $ {{\bar \alpha }_0} = \sqrt {{{10} \mathord{\left/{\vphantom {{10} {11}}} \right. \kern-\nulldelimiterspace} {11}}}  $ and $ {\alpha _l} \sim \mathcal{CN} (0,{1 \mathord{\left/ {\vphantom {1 {11}}} \right. \kern-\nulldelimiterspace} {11}}) $, respectively.
Meanwhile, the distance between the BS and the UE and all the angles are generated as $ {\mathcal{U}} (10,500) $ and ${\mathcal{U}} (-1,1)$, respectively.
In addition, the NMSE is defined as
$ {\rm  {\bf {NMSE} }} = \mathbb{E} \left\{ {{{{{\| {{\bf{\hat H}} - {\bf{H}}} \|}^2}}}/{{{{\left\| {\bf{H}} \right\|}^2}}}} \right\}, $
where $ {\bf{H}} $ is the hybrid-field channel and $ {\bf{\hat H}} $ is the estimation of the channel $ {\bf{H}} $.
Finally, the SNR for the pilot transmission and the uplink date transmission are defined as $ {\rm{SNR}} = p/{\sigma ^2} $ and $ {\rm{SNR}} = p/{\sigma_u ^2} $, respectively.

\subsection{NMSE Performance}

\begin{figure}
  \centering
    \setlength{\abovecaptionskip}{0.cm}
  \includegraphics[width=2.7in]{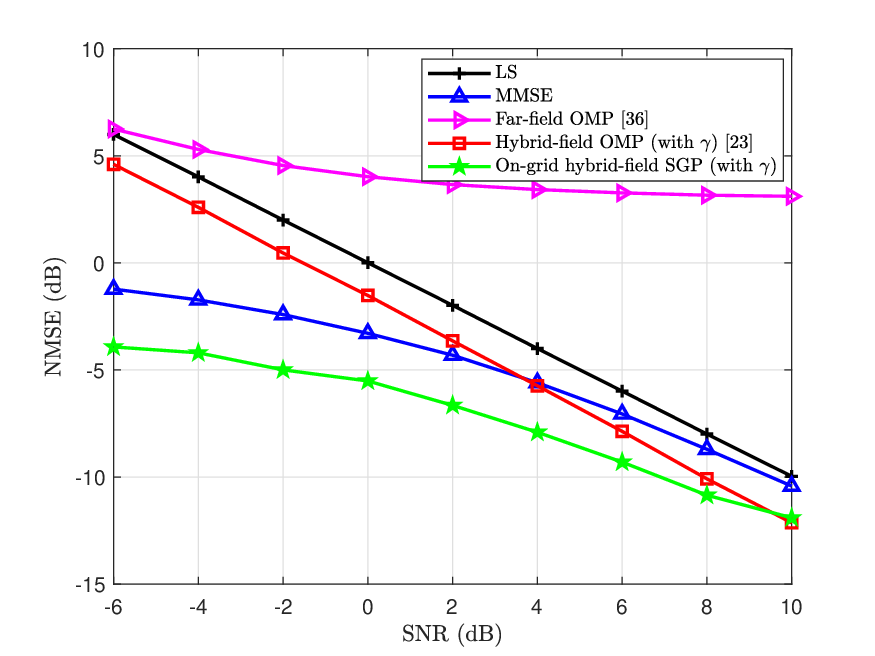}
  \caption{The NMSE performance comparison of the on-grid hybrid-field SGP with the conventional schemes for the scenario with $\gamma$.}\label{SNR-1}
\end{figure}

Fig. \ref{SNR-1} shows a comparison of NMSE performance between the on-grid hybrid-field SGP with the hybrid-field OMP in \cite{[17]} and conventional schemes in the scenario with $\gamma$, where $ \gamma = 0.5 $ and $ \mu = 0.4 $.
The number of antennas at the BS and the UE are set as $ N_b = 256 $ and $ N_u = 1 $, which corresponds to $ M_b = 381 $ and $ M_u = 1 $  sampled grids, respectively.
It can be observed that the proposed on-grid hybrid-field SGP provides the lowest NMSE value than that of the other three considered schemes, especially at low SNRs.
Specifically, at the SNRs of $ -6 $ dB and $0$ dB, the performance gaps between the on-grid hybrid-field SGP and the hybrid-field OMP with $\gamma$ are $8.53$ dB and $ 3.99 $ dB, respectively.
This can be attributed to the on-grid hybrid-field SGP employing the LMS algorithm for channel estimation, in contrast to the LS algorithm employed in the hybrid-field OMP with $\gamma$, leading to enhanced noise resistance.
Moreover, the on-grid hybrid-field SGP exhibits significantly better NMSE performance compared to the far-field OMP, with an improvement of $10.18$ dB and $9.54$ dB at SNRs of $-6$ dB and $0$ dB, respectively.
The reason is that the far-field OMP focuses solely on the characteristics of far-field channels, resulting in a mismatch with the structural characteristics of practical hybrid-field channels.
Indeed, the on-grid hybrid-field SGP significantly outperforms the hybrid-field OMP in low SNR scenarios due to the employment of the LMS algorithm.
Besides, since the on-grid hybrid-field SGP and the hybrid-field OMP adopt the LMS and the LS, respectively, after the pursuing process, these processes both behave similarly as MMSE that result in a reduced performance gap between the on-grid hybrid-field SGP and the hybrid-field OMP.
Overall, the on-grid hybrid-field SGP provides the highest estimation accuracy compared to the other three considered schemes.

\begin{figure}
  \centering
    \setlength{\abovecaptionskip}{0.cm}
  \includegraphics[width=2.7in]{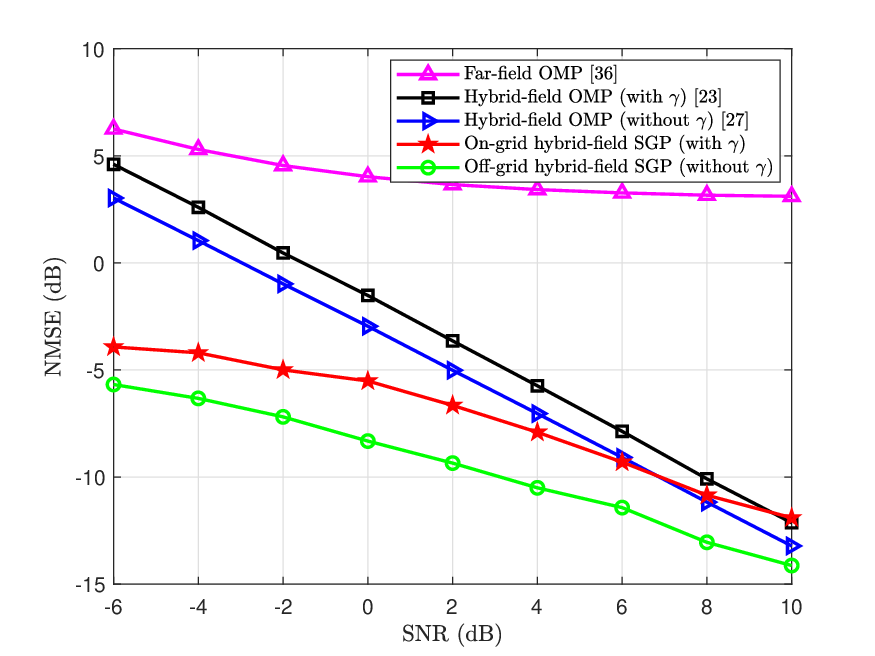}
  \caption{The NMSE performance comparison of the on-grid and off-grid hybrid-field SGP with the hybrid-field OMP for the scenarios with $\gamma$ and without $\gamma$.}\label{SNR-2}
\end{figure}

Fig. \ref{SNR-2} depicts the NMSE performance comparison between the on-grid and off-grid hybrid-field SGP and the hybrid-field OMP across various SNR scenarios, where $ \gamma = 0.5 $, $ \mu = 0.4 $, and $ \nu = 0.03 $.
The first observation is that the on-grid hybrid-field SGP outperforms the hybrid-field OMP with $\gamma$, which aligns with the observation in Fig. \ref{SNR-1}.
However, it is worth noting that since the on-grid hybrid-field SGP does not employ the strategy of selecting the minimal residual vector during the iterative matching pursuit process, it exhibits slightly lower accuracy in identifying the most suitable supports.
Consequently, this results in a lower NMSE performance compared to the off-grid hybrid-field SGP.
On the other hand, although the hybrid-field OMP without $\gamma$ in \cite{[19]} can iteratively adjust the parameter $\gamma$ to determine the minimal residual vector and outperforms the hybrid-field OMP with $\gamma$, the hybrid-field OMP without $\gamma$ still provides a lower NMSE performance than the off-grid hybrid-field SGP.
Indeed, the primary reason for this performance gap is the employment of the LMS algorithm to minimize the mean square error between the hybrid-field channel and its estimation and the alternating minimization approach to refine channel parameters.
All these results show that both the on-grid hybrid-field SGP and the off-grid hybrid-field SGP can significantly enhance the estimation accuracy compared to the hybrid-field OMP with $\gamma$ in \cite{[17]} and the hybrid-field OMP without $\gamma$ in \cite{[19]}.



\begin{figure}
\centering
\subfigure[The NMSE performance comparison against $ N_b $.] {\includegraphics[width=2.7in]{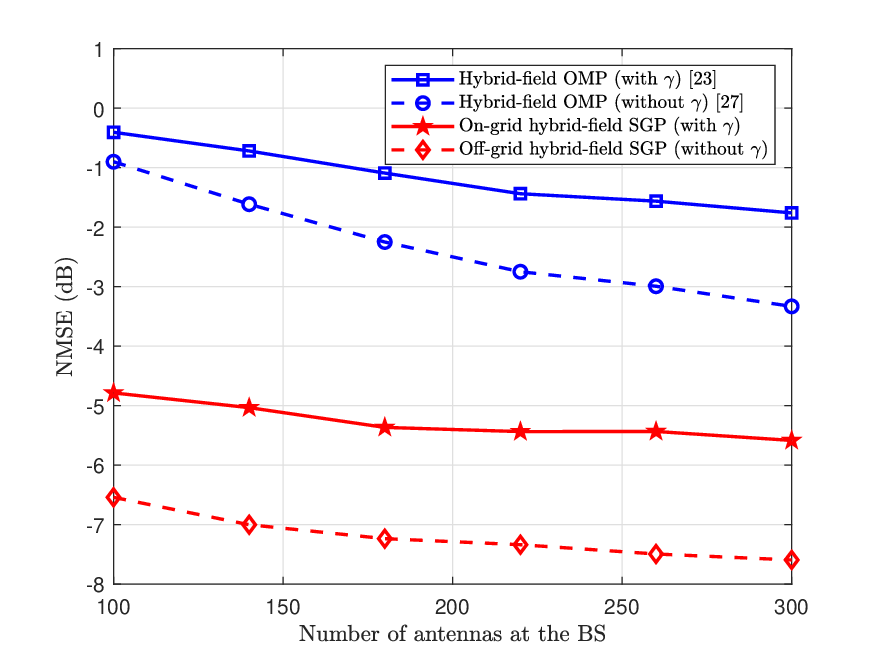}}
\subfigure[The NMSE performance comparison against $ N_u $.] {\includegraphics[width=2.7in]{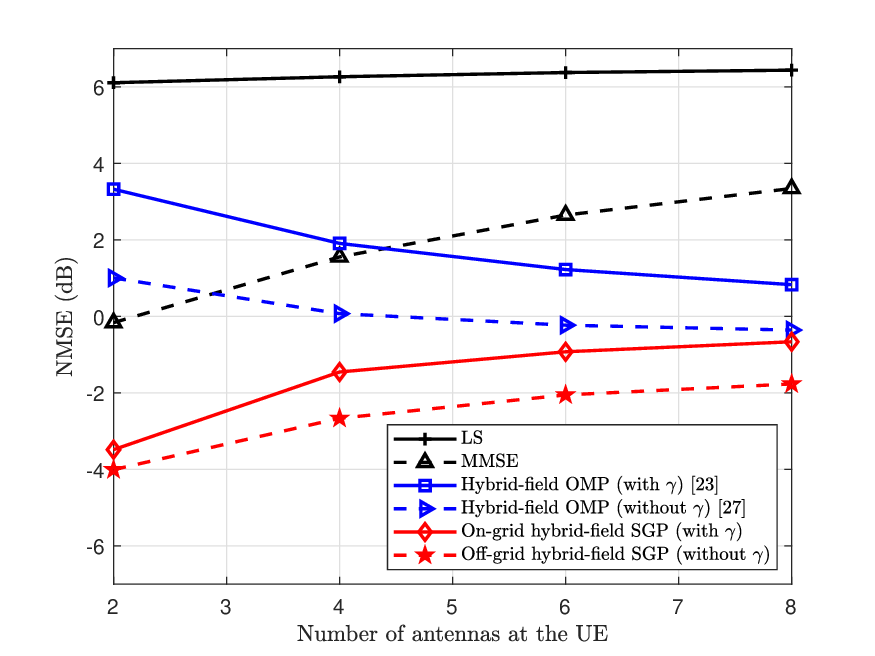}}
\caption{The NMSE performance comparison of the on-grid and off-grid hybrid-field SGP with the existing schemes as a function of (a) the number of antennas at the BS $ N_b $ and (b) the number of antennas at the UE $ N_u $. }
\label{N}
\end{figure}

In Fig. \ref{N} (a), we plot the NMSE performance comparison of the on-grid and off-grid hybrid-field SGP and the hybrid-field OMP as a function of the number of antennas at the BS ($ N_b $), where the parameters are set as follows: $ N_u = 1 $, $ \gamma = 0.5 $, $ \mu = 0.4 $, and $ \nu = 0.08 $, with a SNR of $ 0 $ dB.
As $ N_b $ increases, the performance of all the considered schemes also increases.
This can be attributed to the fact that an increase in $ N_b $ leads to a higher SNR at the BS. As a result, the considered schemes can operate with improved performance and accuracy.
Moreover, with $N_b = 300$, the proposed off-grid hybrid-field SGP demonstrates an approximate NMSE performance improvement of $ 2.00 $ dB, $ 4.26 $ dB, and $ 5.83 $ dB compared to the on-grid hybrid-field SGP, the hybrid-field OMP without $\gamma$, and the hybrid-field OMP with $\gamma$, respectively.
In addition, it is noteworthy that the performance gaps between the off-grid hybrid-field SGP and the on-grid hybrid-field SGP, as well as between the hybrid-field OMP without $\gamma$ and the hybrid-field OMP with $\gamma$, initially increase rapidly with the growth of $ N_b $ but with diminishing returns.
This indicates that their performance gains gradually reduce as $ N_b $ increases.
The reason for this trend can be that the increment in the number of antennas at the BS results in a higher-dimensional received signal.
Consequently, channel estimation processes need to handle a greater number of parameters and the considered channel estimation schemes become less effective when dealing with such high-dimensional challenges.



In Fig. \ref{N} (b), we compare the NMSE performance of the on-grid and off-grid hybrid-field SGP with the existing schemes against the number of antennas at the UE ($ N_u $), where $ N_b = 256 $, $ \gamma = 0.5 $, $ \mu = 0.4 $, and $ \nu = 0.07 $, with the SNR of $ -6 $ dB.
The first observation is that with the increase of $ N_u $, the NMSE performance of the hybrid-field OMP without $\gamma$ and the hybrid-field OMP with $\gamma$ both increase, while the NMSE performance of the off-grid hybrid-field SGP and the on-grid hybrid-field SGP both decrease.
The rationale behind this phenomenon is that as $ N_u $ increases, there is an inevitable rise in channel sparsity.
However, an increase in $ N_u $ leads to a decrease in the performance of MMSE, while the performance of LS remains unchanged.
Consequently, the performance of both the hybrid-field OMP without $\gamma$ and with $\gamma$ is expected to improve.
In contrast, the performance of the on-grid and off-grid hybrid-field SGP experience a decline.
Nevertheless, it is important to note that the off-grid hybrid-field SGP consistently outperforms the other considered channel estimation schemes and maintains the best NMSE performance across the different values of $ N_u $.


\begin{figure}
\centering
\subfigure[The NMSE performance against $\gamma$.]{\includegraphics[width=2.7in]{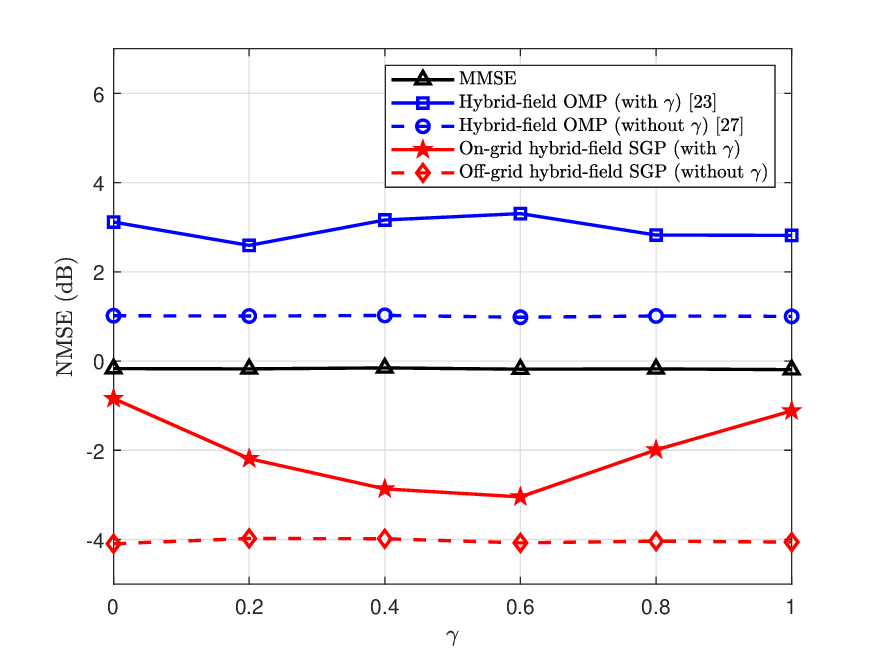}}
\subfigure[The NMSE performance against $ L $.]{\includegraphics[width=2.7in]{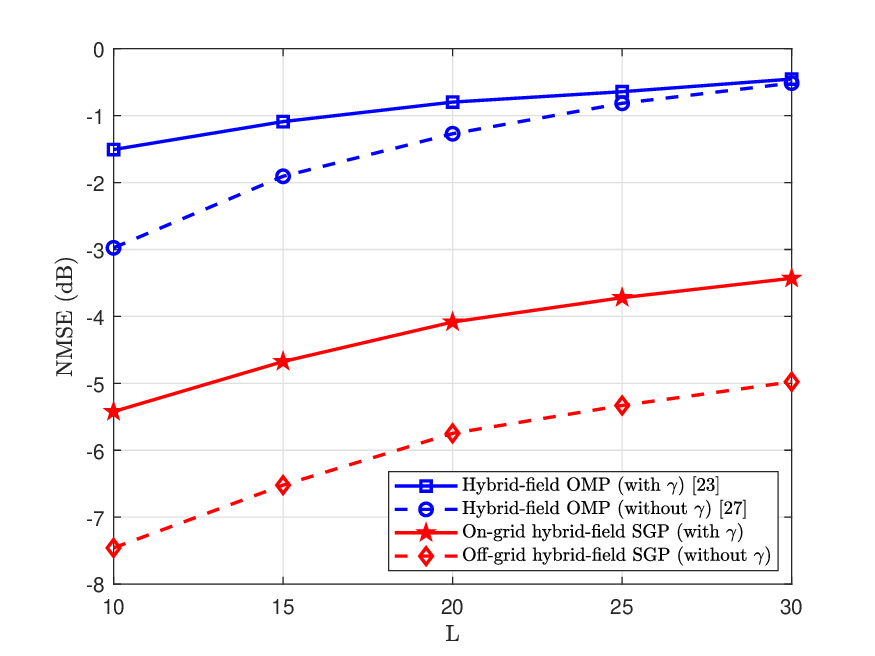}}
\caption{The NMSE performance comparison of the on-grid and off-grid hybrid-field SGP with the existing schemes as a function of (a) $\gamma$ and (b) $L$.}
\label{gamma-L}
\end{figure}

Fig. \ref{gamma-L} (a) compares the NMSE performance of the on-grid and off-grid hybrid-field SGP and the hybrid-field OMP as a function of $ \gamma $ for the scenarios with $\gamma$ and without $\gamma$.
Here, $ N_b = 256 $, $ N_u = 2 $, $ \mu = 0.1 $, and $ \nu = 0.07 $, with the SNR of $ -6 $ dB.
We can find that the proposed off-grid hybrid-field SGP consistently achieves the lowest NMSE value across all values of the parameter $\gamma$.
Specifically, the off-grid hybrid-field SGP provides an average NMSE performance gain of $ 3.89 $ dB and $ 4.92 $ dB compared to MMSE and the hybrid-field OMP without $\gamma$.
This once again verifies the effectiveness of the off-grid hybrid-field SGP in the hybrid-field channel estimation.
In addition, the off-grid hybrid-field SGP outperforms both the on-grid hybrid-field SGP and the hybrid-field OMP with $\gamma$ for all ranges of $\gamma$.
Notably, in scenarios where $\gamma = 1$ and $\gamma = 0$ (representing entirely far-field and near-field scenarios, respectively), the off-grid hybrid-field SGP exhibits a significantly superior NMSE performance.
This notable improvement can be attributed to potential misalignment between the distance and angle parameters of each path and the sampled grids of the angular-domain or the polar-domain transformation matrices.
This misalignment occurs due to the limited dimensions of these matrices and can lead to NMSE degradation for the on-grid hybrid-field SGP and the hybrid-field OMP with $\gamma$.
However, the proposed off-grid hybrid-field SGP can iteratively adjust the value of $\gamma$ to identify the most matched polar-domain and angular-domain supports and then refine channel parameters, enabling effective recovery of the channel with improved estimation performance.

%

Fig. \ref{gamma-L} (b) shows a comparison of the NMSE performance as a function of the number of channel paths $ L $, with the following parameters: $ N_b = 256 $, $ N_u = 1 $, $ \gamma = 0.5 $, $ \mu = 0.8 $, $ \nu = 0.08 $, and a SNR of $ 0 $ dB.
Initially, it is evident that the NMSE performance for the considered schemes declines as $L$ increases.
This trend is attributed to the diminishing channel sparsity with more channel paths, adversely affecting the performance of CS-based approaches.
Furthermore, the performance gap between the hybrid-field OMP with $\gamma$ and the hybrid-field OMP without $\gamma$ narrows with an increase in $L$.
Conversely, the performance gap between the on-grid hybrid-field SGP and the off-grid hybrid-field SGP remains constant.
This phenomenon suggests that the advantage of exploring all potential $\gamma$ values diminishes with more channel paths, whereas the off-grid hybrid-field SGP continues to achieve notable performance enhancements through iterative channel parameter refinement using the alternating minimization approach.
Overall, the simulation outcomes underscore the robustness of the proposed algorithms against various factors, including noise levels, number of antennas, the proportion of far and near field paths $\gamma$, and the number of channel paths.

\subsection{Achievable Rate}\label{The SE performance}



\begin{figure}
\centering
\subfigure[The achievable rates against the SNR.] {\includegraphics[width=2.7in]{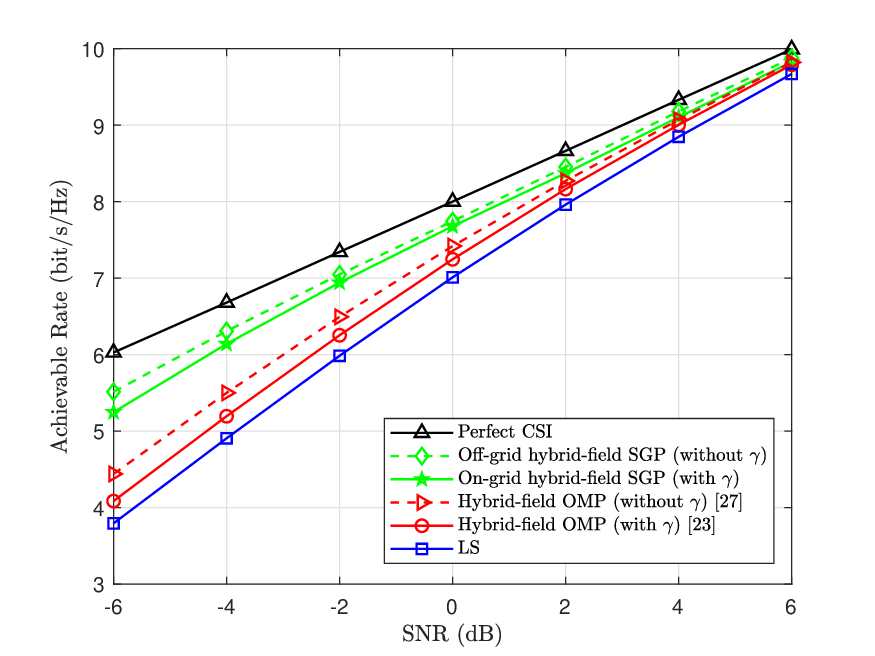}}
\subfigure[The achievable rates against the number of antennas at the BS.] {\includegraphics[width=2.7in]{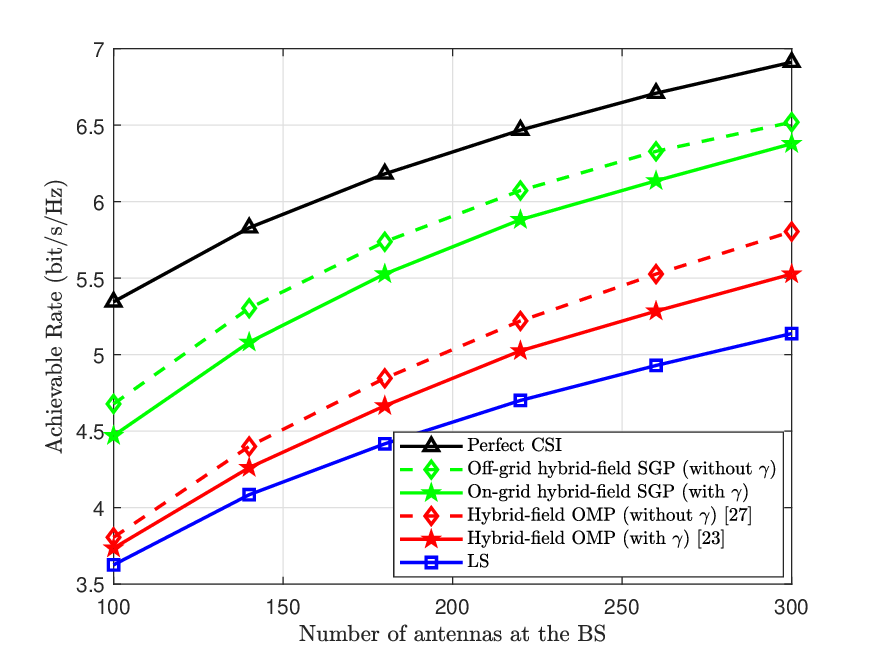}}
\caption{The achievable rates of LS, the hybrid-field OMP, the on-grid and off-grid hybrid-field SGP, and the perfect CSI against (a) the SNR and (b) the number of antennas at the BS.}
\label{SE}
\end{figure}	


Fig. \ref{SE} (a) compares the achievable rates of LS, the hybrid-field OMP, the on-grid and off-grid hybrid-field SGP, and the perfect CSI against the SNR.
The following parameters are adopted: $ N_b = 256 $, $ N_u = 1 $, $ \gamma = 0.4 $, $ \mu = 0.4 $, $ \nu = 0.07 $.
The result of the perfect CSI is based on setting $ {\bf{ V}} = {\bf{ H}} $.
The first observation is that as the SNR increases, the achievable rates for all the schemes improve, while the performance gaps between the achievable rates for the perfect CSI and those of all the channel estimation schemes become narrow.
This can be attributed to two factors: first, increasing the SNR can increase the expected signal power, leading to improved achievable rates.
Second, as the SNR increases, the estimation accuracy of all the channel estimation schemes is enhanced, mitigating the adverse impact of imperfect CSI on the achievable rates.
Moreover, the proposed off-grid hybrid-field SGP consistently outperforms other channel estimation schemes in terms of achievable rates.
This superiority can be attributed to the excellent NMSE performance achieved by the off-grid hybrid-field SGP, resulting in a more accurate acquisition of CSI.
Therefore, the BS can utilize more precise estimated channel information when processing signals from the BS, leading to higher achievable rates.
Fig. \ref{SE} (b) provides a comprehensive comparison of the achievable rates against the number of antennas at the BS.
These results are presented for the following parameters: $ N_u = 1 $, $ \gamma = 0.3 $, $ \mu = 0.4 $, $ \nu = 0.08 $, and a SNR of $ -4 $ dB.
It is evident that the achievable rates for all the schemes increase as the number of antennas at the BS, denoted as $N_b$, increases.
This growth can be attributed to the rise in the SNR at the BS with an increase in $N_b$.
As a result, the considered schemes can operate with improved NMSE performance, leading to higher achievable rates.
Moreover, the proposed hybrid-field SGP can achieve higher achievable rates compared to other channel estimation schemes, irrespective of the scenarios with $\gamma$ or without $\gamma$.
\section{Conclusions}\label{Conclusions}

In this work, we proposed two novel hybrid-field channel estimation schemes, namely the on-grid hybrid-field SGP and the off-grid hybrid-field SGP, aimed at enhancing the precision of the hybrid-field channel estimation within XL-MIMO systems with a multi-antenna UE.
It is clear that the employment of the LMS algorithm in both the on-grid and off-grid hybrid-field SGP can significantly improve the quality of the hybrid-field channel estimation compared to the hybrid-field OMP with $\gamma$ and the hybrid-field OMP without $\gamma$.
This improvement is particularly pronounced at lower SNR values.
Through simulations, we demonstrated that the proposed schemes can provide superior NMSE performance, while simultaneously offering reduced computational complexity in the hybrid-field channel estimation of XL-MIMO systems.
What is even more interesting is that the off-grid hybrid-field SGP can effectively keep a satisfactory NMSE performance, irrespective of the values of $\gamma$. 
As for future works, we will further investigate the impact of the spatial non-stationary characteristics and the EM polarization property of near-field channels on the estimation performance for XL-MIMO systems.

\section*{Appendix}

This appendix derive the explicit derivation of $ \nabla {\mathcal L}({\bf{\hat r}},{{{\bm{\hat \theta }}}})$ with respect to $\frac{1}{{{\bf{\hat r}}}} = [\frac{1}{{{{\hat r}_{{\rm{R}},1}}}},\frac{1}{{{{\hat r}_{{\rm{R}},2}}}}, \cdots ,\frac{1}{{{{\hat r}_{{\rm{R}},{L_{\rm{N}}}}}}},\frac{1}{{{{\hat r}_{{\rm{T}},1}}}},\frac{1}{{{{\hat r}_{{\rm{T}},2}}}}, \cdots ,\frac{1}{{{{\hat r}_{{\rm{T}},{L_{\rm{N}}}}}}}]^T$ and $ {{\bm{\hat \theta }}} = [ {{{\bm{\hat \theta }}}_{{\rm{R}},{\rm{near}}}}^T, {{{\bm{\hat \theta }}}_{{\rm{T}},{\rm{near}}}}^T, {{{\bm{\hat \theta }}}_{{\rm{R}},{\rm{far}}}}^T, {{{\bm{\hat \theta }}}_{{\rm{T}},{\rm{far}}}}^T  ]^T $.
For any variable $ z \in \{ \frac{1}{{{\bf{\hat r}}}}, {{\bm{\hat \theta }}} \} $, the gradient of $  {\mathcal L}({\bf{\hat r}},{{{\bm{\hat \theta }}}})$ can be computed as
\begin{equation}\label{(a1)}
\begin{aligned}
\frac{{\partial {\mathcal L}\left( {{\bf{\hat r}},{\bm{\hat \theta }}} \right)}}{{\partial z}} =  - {\rm{tr}}\left\{ {{\bf{y}}_t^H\frac{{\partial {\bf{\Psi }}}}{{\partial z}}{{\bf{y}}_t}} \right\}.
\end{aligned}
\end{equation}
Since $ {\bf{\Psi }} = {\bf{\hat A}}{{{\bf{\hat A}}}^\dag } $ and $ {{{\bf{\hat A}}}^\dag } = {({{{\bf{\hat A}}}^H}{\bf{\hat A}})^{ - 1}}{{{\bf{\hat A}}}^H} $, we have
\begin{equation}\label{(a2)}
\begin{aligned}
\frac{{\partial {\bf{\Psi }}}}{{\partial z}} &= \frac{{\partial {\bf{\hat A}}}}{{\partial z}}{( {{{{\bf{\hat A}}}^H}{\bf{\hat A}}} )^{ - 1}}{{{\bf{\hat A}}}^H} + {\bf{\hat A}}\frac{{\partial {{( {{{{\bf{\hat A}}}^H}{\bf{\hat A}}} )}^{ - 1}}}}{{\partial z}}{{{\bf{\hat A}}}^H} \\ &+ {\bf{\hat A}}{( {{{{\bf{\hat A}}}^H}{\bf{\hat A}}} )^{ - 1}}\frac{{\partial {{{\bf{\hat A}}}^H}}}{{\partial z}}.
\end{aligned}
\end{equation}
According to the inverse matrix differentiation law that $ d{{\bf{A}}^{ - 1}} =  - {{\bf{A}}^{ - 1}}\left( {d{\bf{A}}} \right){{\bf{A}}^{ - 1}} $, the gradient of $ {{{( {{{{\bf{\hat A}}}^H}{\bf{\hat A}}} )}^{ - 1}}} $ is
\begin{equation}\label{(a3)}
\begin{aligned}
\frac{{\partial {{( {{{{\bf{\hat A}}}^H}{\bf{\hat A}}} )}^{ - 1}}}}{{\partial z}} =  &- {( {{{{\bf{\hat A}}}^H}{\bf{\hat A}}} )^{ - 1}}\left( {\frac{{\partial {{{\bf{\hat A}}}^H}}}{{\partial z}}{\bf{\hat A}} + {{{\bf{\hat A}}}^H}\frac{{\partial {\bf{\hat A}}}}{{\partial z}}} \right)  \\
& \times {( {{{{\bf{\hat A}}}^H}{\bf{\hat A}}} )^{ - 1}}.
\end{aligned}
\end{equation}
Therefore, the remaining task is to derive the gradient of $ {{\bf{\hat A}}} $ with respect to $ {{{\bm{\hat \theta }}}_{{\rm{R}},{\rm{far}}}} $, $ {{{\bm{\hat \theta }}}_{{\rm{T}},{\rm{far}}}} $, $ {{{\bm{\hat \theta }}}_{{\rm{T}},{\rm{near}}}} $, $ {{{\bm{\hat \theta }}}_{{\rm{R}},{\rm{near}}}} $, $  \frac{1}{{{{\bf{\hat r}}}_{{\rm{R}},{\rm{near}}}}} $, and $ \frac{1}{{{{\bf{\hat r}}}_{{\rm{T}},{\rm{near}}}}} $.

For the far-field angle $ {{\hat \theta }_{{\rm{T}},{l_{\rm{f}}}}} \in   [{{\hat \theta }_{{\rm{T}},1}},{{\hat \theta }_{{\rm{T}},2}}, \cdots ,{{\hat \theta }_{{\rm{T}},{L_{\rm{F}}}}}]^T $, we have
\begin{equation}\label{(a4)}
\begin{aligned}
\frac{{\partial {\bf{\hat A}}}}{{\partial {{\hat \theta }_{{\rm{T}},{l_{\rm{f}}}}}}} = [ {0, \cdots ,0,\sqrt {\frac{{\tau p}}{{{N_u}}}} \frac{{\partial {{\bf{a}}^*}({{\hat \theta }_{{\rm{T}},{l_{\rm{f}}}}})}}{{\partial {{\hat \theta }_{{\rm{T}},{l_{\rm{f}}}}}}} \otimes {\bf{a}}({{\hat \theta }_{{\rm{R}},{l_{\rm{f}}}}}),0, \cdots ,0} ],
\end{aligned}
\end{equation}
and the $n_u$-th entry of $ {\partial {{\bf{a}}^*}({{\hat \theta }_{{\rm{T}},{l_{\rm{f}}}}})} / {\partial {{\hat \theta }_{{\rm{T}},{l_{\rm{f}}}}}} $ is $- j({n_u} - 1)\pi {e^{ - j({n_u} - 1)\pi {{\hat \theta }_{{\rm{T}},{l_{\rm{f}}}}}}}/ \sqrt{N_u}$.
For the far-field angle $ {{\hat \theta }_{{\rm{R}},{l_{\rm{f}}}}} \in [{{\hat \theta }_{{\rm{R}},1}},{{\hat \theta }_{{\rm{R}},2}}, \cdots ,{{\hat \theta }_{{\rm{R}},{L_{\rm{F}}}}}]^T $, we have
\begin{equation}\label{(a6)}
\begin{aligned}
\frac{{\partial {\bf{\hat A}}}}{{\partial {{\hat \theta }_{{\rm{R}},{l_{\rm{f}}}}}}} = \left[ {0, \cdots ,0,\sqrt {\frac{{\tau p}}{{{N_u}}}} {{\bf{a}}^*}({{\hat \theta }_{{\rm{T}},{l_{\rm{f}}}}}) \otimes \frac{{\partial {\bf{a}}({{\hat \theta }_{{\rm{R}},{l_{\rm{f}}}}})}}{{\partial {{\hat \theta }_{{\rm{R}},{l_{\rm{f}}}}}}},0, \cdots ,0} \right].
\end{aligned}
\end{equation}
For the near-field angle $ {{\hat \theta }_{{\rm{T}},{l_{\rm{n}}}}} \in [{{\hat \theta }_{{\rm{T}},1}},{{\hat \theta }_{{\rm{T}},2}}, \cdots ,{{\hat \theta }_{{\rm{T}},{L_{\rm{N}}}}}]^T $, we have
\begin{equation}\label{(a7)}
\begin{aligned}
\frac{{\partial {\bf{\hat A}}}}{{\partial {{\hat \theta }_{{\rm{T}},{l_n}}}}} &= \sqrt {\frac{{\tau p}}{{{N_u}}}} [  0,\cdots ,0,\\
&\frac{{\partial {{\bf{a}}^*}({{\hat r}_{{\rm{T}},{l_n}}},{{\hat \theta }_{{\rm{T}},{l_n}}})}}{{\partial {{\hat \theta }_{{\rm{T}},{l_n}}}}} \otimes {\bf{a}}({{\hat r}_{{\rm{R}},{l_n}}},{{\hat \theta }_{{\rm{R}},{l_n}}}),0, \cdots  ,0 ],
\end{aligned}
\end{equation}
and the $n_u$-th entry of $ {\partial {{\bf{a}}^*}({{\hat r}_{{\rm{T}},{l_n}}},{{\hat \theta }_{{\rm{T}},{l_n}}})}/{\partial {{\hat \theta }_{{\rm{T}},{l_n}}}} $ is $ j\frac{{2\pi }}{\lambda }\frac{{2{n_u} - {N_u} - 1}}{2}\frac{\lambda }{2}\frac{1}{{\sqrt {{N_u}} }}{e^{j\frac{{2\pi }}{\lambda }\left( {r_{{\rm{T}},{l_{\rm{n}}}}^{\left( {{n_u}} \right)} - {r_{{\rm{T}},{l_{\rm{n}}}}}} \right)}}  $.
For the near-field distance $ \frac{1}{{{{\hat r}_{{\rm{T}},{l_n}}}}} \in [\frac{1}{{{{\hat r}_{{\rm{T}},1}}}},\frac{1}{{{{\hat r}_{{\rm{T}},2}}}}, \cdots ,\frac{1}{{{{\hat r}_{{\rm{T}},{L_{\rm{N}}}}}}}]^T $, we have
\begin{equation}\label{(a8)}
\begin{aligned}
\frac{{\partial {\bf{\hat A}}}}{{\partial \frac{1}{{{{\hat r}_{{\rm{T}},{l_n}}}}}}}&= \sqrt {\frac{{\tau p}}{{{N_u}}}} [ 0, \cdots ,0,\\
&\frac{{\partial {{\bf{a}}^*}({{\hat r}_{{\rm{T}},{l_n}}},{{\hat \theta }_{{\rm{T}},{l_n}}})}}{{\partial \frac{1}{{{{\hat r}_{{\rm{T}},{l_n}}}}}}} \otimes {\bf{a}}({{\hat r}_{{\rm{R}},{l_n}}},{{\hat \theta }_{{\rm{R}},{l_n}}}),0, \cdots ,0 ],
\end{aligned}
\end{equation}
and the $n_u$-th entry of $ {\partial {{\bf{a}}^*}({{\hat r}_{{\rm{T}},{l_n}}},{{\hat \theta }_{{\rm{T}},{l_n}}})}/{\partial \left(\frac{1}{{{{\hat r}_{{\rm{T}},{l_n}}}}}\right)} $ is $ j\frac{{2\pi }}{\lambda }\left( { - {{\left( {\frac{{2{n_u} - {N_u} - 1}}{2}\frac{\lambda }{2}} \right)}^2}} \right)\frac{1}{{\sqrt {{N_u}} }}{e^{j\frac{{2\pi }}{\lambda }\left( {r_{{\rm{T}},{l_{\rm{n}}}}^{\left( {{n_u}} \right)} - {r_{{\rm{T}},{l_{\rm{n}}}}}} \right)}} $. As for the near-field angle $ {{\hat \theta }_{{\rm{R}},{l_{\rm{n}}}}} \in [{{\hat \theta }_{{\rm{R}},1}},{{\hat \theta }_{{\rm{R}},2}}, \cdots ,{{\hat \theta }_{{\rm{R}},{L_{\rm{N}}}}}]^T $ and distance $ \frac{1}{{{{\hat r}_{{\rm{R}},{l_n}}}}} \in [\frac{1}{{{{\hat r}_{{\rm{R}},1}}}},\frac{1}{{{{\hat r}_{{\rm{R}},2}}}}, \cdots ,\frac{1}{{{{\hat r}_{{\rm{R}},{L_{\rm{N}}}}}}}]^T $, their gradients are similar to \eqref{(a7)} and \eqref{(a8)}, respectively. Finally, collecting all of the angle and distance terms in a column vector, we can obtain the gradient $ {\nabla _{{\bm{\hat \theta }}}}{\mathcal L}({{{\bf{\hat r}}}^{n - 1}},{\bm{\hat \theta }}) $ and $ {\nabla _{\frac{1}{{{\bf{\hat r}}}}}}{\mathcal L}({\bf{\hat r}},{{{\bm{\hat \theta }}}^n}) $.

\bibliographystyle{IEEEtran}
\bibliography{IEEEabrv,Ref}

\end{document}